\documentclass[iop]{emulateapj} 

\usepackage{natbib}
\usepackage{graphicx}
\usepackage{amsmath}
\usepackage{color}
\newcommand{\Msun}{M_{\odot}}

\newcommand{\fesc}{f_{\rm esc}}

\newcommand{\art}{\rm ART^{2}}
\newcommand{\A}{\rm \AA}
\newcommand{\La}{L_{\rm{Ly\alpha}}}

\newcommand{\EW}{\rm EW}

\newcommand{\Mpc}{\rm {Mpc}}

\newcommand{\Msunyr}{\rm {M_{\odot}\; yr^{-1}}}
\newcommand{\fescalpha}{f_{\rm esc}^{\lya}}

\newcommand{\ergs}{{\rm erg~s^{-1}}}

\newcommand{\lya}{\ifmmode {\rm Ly}\alpha \else Ly$\alpha$\fi}
\def\msunyr{\ifmmode M_{\odot} {\rm yr}^{-1} \else M$_{\odot}$ yr$^{-1}$\fi}

\begin{document}                          
%
%

\title{Cold Accretion In Early Galaxy Formation And Its $\lya$ Signatures}    

%
%
\author
{
Hidenobu Yajima\altaffilmark{1,2},
Yuexing Li\altaffilmark{3,4},
Qirong Zhu\altaffilmark{3,4}, 
Tom Abel\altaffilmark{5}
}

\affil{$^{1}$ Department of Earth \& Space Science, Graduate School of Science, Osaka University, 1-1 Machikaneyama, 
   Toyonaka, Osaka 560-0043, Japan}

\affil{$^{2}$ Institute for Astronomy, University of Edinburgh, Royal Observatory, Edinburgh, EH9 3HJ, UK}

\affil{$^{3}$Department of Astronomy \& Astrophysics, The Pennsylvania State University, 
525 Davey Lab, University Park, PA 16802, USA}

\affil{$^{4}$Institute for Gravitation and the Cosmos, The Pennsylvania State University, University Park, PA 16802, USA}

\affil{$^{5}$Kavli Institute for Particle Astrophysics and Cosmology, SLAC National Accelerator Laboratory, Stanford University, \\
2575 Sand Hill Road, Menlo Park, CA 94025, USA}

\email{yajima@roe.ac.uk}

%
%
\begin{abstract}

The $\lya$ emission has played an important role in detecting high-redshift galaxies, 
including recently distant ones at redshift $z > 7$.
It may also contain important information on the origin of these galaxies. Here, we investigate the formation of a typical $L^*$ galaxy and its observational signatures at the earliest stage, by combining a cosmological hydrodynamic simulation with three-dimensional radiative transfer calculations using the newly improved $\art$ code. Our cosmological simulation uses the Aquila initial condition which zooms in onto a Milky Way-like halo with high resolutions, and our radiative transfer couples multi-wavelength continuum, $\lya$ line, and ionization of hydrogen. We find that the modeled galaxy starts to form at redshift $z \sim 24$ through efficient accretion of cold gas, which produces a strong $\lya$ line with a luminosity of $\La \sim 10^{42}~\ergs$ as early as $z \sim 14$. The $\lya$ emission appears to trace the cold, dense gas. The lines exhibit asymmetric, {single}-peak profiles, and are shifted to the blue wing, a characteristic feature of gas inflow. Moreover, the contribution to the total $\lya$ luminosity by excitation cooling increases with redshift, and it becomes dominant at $z \gtrsim 6$. We predict that $L^*$ galaxies such as the modeled one may be detected at $z \lesssim 8$ by JWST and ALMA with a reasonable integration time. 
Beyond redshift 12, however, only $\lya$ line may be observable by spectroscopic surveys. 
Our results suggest that $\lya$ line is one of the most powerful tools to detect the first generation of galaxies, and to decipher their formation mechanism.

\end{abstract}

%
%

%
%
\section{Introduction}

The quest for the first galaxies formed at the cosmic dawn is a major frontier in both observational and theoretical cosmology \citep{Bromm2011}. Over the past few years, significant progress has been made in detecting galaxies at redshift $z \gtrsim 6$, using either broad-band colors \citep[e.g.,][]{Bouwens2004, Bouwens2006, Bouwens2010, Bouwens2011A}, or narrow-band $\lya$ emission line \citep[e.g.,][]{Malhotra04, Iye06, Stark07, Hu2010, Lehnert10, Vanzella11, Stark2011, Ono11, Kashikawa2011, Shibuya11, Finkelstein13}. In particular, the $\lya$ line has played an important role in identifying and confirming distant galaxies, the so-called $\lya$ emitters (LAEs), including currently the record holder at $z=8.6$ \citep{Lehnert10}. These remarkable observations indicate that galaxies formed less than a few hundred million years after the Big Bang. 

Despite the rapidly increasing number of detections, the origin and nature of these distant galaxies, however, remain open questions \citep{Bromm2009}. Recently, a number of state-of-the-art simulations have started to address this issue \citep[e.g.,][]{Wise2007, Wise2008A, Wise2008B, Greif2010, Wise2010, Jeon2011}. These studies focused on a halo in a small volume (1 Mpc) with high resolutions, and suggested that the formation of the first galaxies is closely tied to the formation of the first stars (so-called Pop~III stars) and the feedback from them, and that these galaxies likely consist of second- or third generation of stars formed from enriched gas, similar to the present-day stars. 

In this work, we explore the physical conditions of early galaxy formation on a larger scale. In particular, we focus on the gas properties and the $\lya$ emission from it. Recent simulations have revealed that a large amount of gas penetrate deep inside dark matter halos as cold, filamentary streams \citep{Katz03, Birnboim03, Keres05, Keres09, Dekel06, Ocvirk08, Brooks09, Dekel09}, and \citet{Dekel09} showed that massive galaxies at $z = 2 - 3$ can actively form stars from inflow of cold gas. More recently, \citet{DiMatteo11} suggested that massive galaxies at $z \gtrsim 6$ can grow by cold accretion and evolve with black holes. Such streams of cold gas may produce a large number of $\lya$ photons via excitation cooling process, and give rise to the $\lya$ emission detected in the early galaxies \citep{Dijkstra09, Faucher10, Latif11, Yajima12b, Yajima12c}. 

We combine a multi-scale cosmological hydrodynamic simulation with
multi-wavelength radiative transfer (RT) calculations. The simulation uses the
Aquila initial condition and follows the formation and evolution of a Milky
Way-size galaxy {\citep{Wadepuhl11, Scannapieco12}}. It covers a large
dynamical range from a $100~h^{-1} \Mpc$ box down to a $\sim 5~h^{-3}\Mpc$
zoom-in region, which is ideal to study the gas inflow on a large scale. The
RT calculations uses the three-dimensional Monte Carlo RT code $\art$ by
\cite{Li08, Yajima12b}. The $\art$ code couples multi-wavelength continuum, $\lya$ line, and ionization of hydrogen, which is critical to study the $\lya$ and multi-band properties of the early galaxies. 

The paper is organized as follows. We describe our cosmological simulation in \S2, and the RT calculations in \S3. In \S4, we present the results, which include the $\lya$ properties, and detectability by upcoming missions {\it James Webb Space Telescope} (JWST)
 and {\it Atacama Large Millimeter Array} (ALMA). We discuss the implications and limitations of our model in \S5, and summarize in \S6. 

%
%
\section{Model \& Methodology}

We carry out a cosmological simulation with the Aquila initial condition which
can reproduce a Milky Way-like galaxy at $z=0$  \citep{Springel08a, Scannapieco12}.
The whole simulation box is $100~h^{-1} \Mpc$ on each side with a zoom-in
region of a size of $5\times 5\times 5~h^{-3}\Mpc^{3}$. The spatial
resolution in the zoom-in region is $\sim 250~h^{-1}$ pc and the mass
resolution is $1.8 \times 10^{6}~ h^{-1} \Msun$ for dark matter particles, $3
\times  10^{5}~ h^{-1} \Msun$ for gas, and  $1.5 \times 10^{5}~ h^{-1} \Msun$
for star particles. The cosmological parameters used in the simulation are
$\Omega_{m }= 0.25$, $\Omega_{\Lambda} = 0.75$, $\sigma_{8} = 0.9$ and
$h=0.73$, consistent with the five-year results of the WMAP
\citep{Komatsu09}. The simulation was performed using the N-body/Smoothed
Particle Hydrodynamics (SPH) code GADGET-3 \citep{Springel01, Springel05e}. The
specifics of the simulation were described in Zhu et al. (2012), and we refer the
readers to that paper for more details.

\label{sec:rt}

In this work, we use the 3D Monte Carlo RT code, All-wavelength Radiative
Transfer with Adaptive Refinement Tree ($\art$) to study the multi-wavelength
properties of the model galaxies. The $\art$ code includes continuum photons from X-ray to radio, $\lya$ line, and ionization structure in 
the adaptive refinement grids.  The detailed prescriptions of the code were
presented in \cite{Li08} and \cite{Yajima12b}.
The $\lya$ emission comes from the recombination and de-excitation process, 
\begin{equation}
\epsilon_{\lya} = f_{\alpha } \alpha_{\rm B} h \nu_{\rm \alpha} n_{\rm e} n_{\rm HII}
+ C_{\rm Ly \alpha} n_{\rm e} n_{\rm HI},
\end{equation}
where $\alpha_{\rm B}$ is the case B recombination coefficient, and $f_{\alpha}$ is the average number of $\lya$ photons produced per case B recombination. The $\alpha_{\rm B}$ derived in \citet{Hui97} is used. Due to the small dependence of $f_{\alpha}$ to temperature, we assume $f_{\alpha} = 0.68$ at everywhere \citep{Osterbrock06}. 
The $C_{\rm Ly \alpha}$ is the collisional excitation coefficient,
$C_{\rm Ly\alpha} = 3.7 \times 10^{-17} {\rm exp}(- h\nu_{\alpha}/kT) T^{-1/2}~\rm erg \; s^{-1}\; cm^{3}$ \citep{Osterbrock06}.
The $\lya$ emissivity and opacity highly depend on the ionization structure in the galaxies. 
We at first calculate the ionization structure due to internal stellar sources, then simulate the $\lya$ RT. We cast $N_{\rm ph} = 10^{5}$ photon packets for each ionizing, $\lya$, and non-ionizing components, which showed good convergence \citep{Yajima12b, Yajima12c}.
In addition, interstellar dust is included to consider the dust extinction of $\lya$ and continuum photons, and to simulate the dust thermal emission \citep[see also][]{Yajima12b}. 
The intrinsic spectral energy distributions (SEDs) of stars are calculated by
GALAXEV \citep{Bruzual03} with the assumption of Salpeter IMF. For the SED of
AGNs, the broken power law is used  \citep{Li08}.
For the intrinsic SEDs, we do not include $\lya$ line as nebula emission. 
This is because we here calculate the radiative transfer of ionizing photons and ionization structure. 
Some fraction of ionizing photons are absorbed 
in situ, and converted to $\lya$ photons via the recombination process. 
Therefore, the nebula emission at $\lya$ line is included in the post-processing calculations.

%
%

\section{Results}

In our previous work, we have presented the formation history of a MW galaxy
in \citet{Zhu12}, and have applied $\art$ to the Aquila simulation to study 
the multi-band properties of the MW progenitors \citep{Yajima12b}, and the
escape of $\lya$ and continuum photons \citep{Yajima12c}. In this paper, we
focus on the  earliest evolutionary stage of the MW and the $\lya$ properties from $z \sim 6 - 14$.

\subsection{The Accretion of Cold Gas}

\begin{figure*}
\begin{center}
\includegraphics[scale=0.37, bb=105 165 510 570, clip=true]{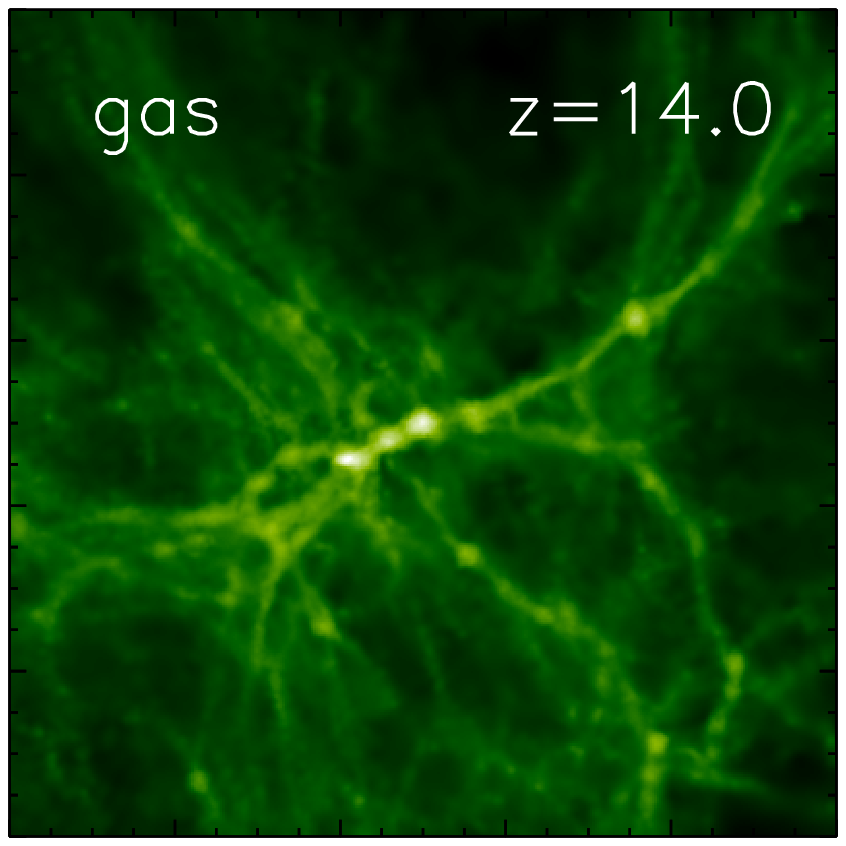}
\includegraphics[scale=0.37, bb=105 165 560 570, clip=true]{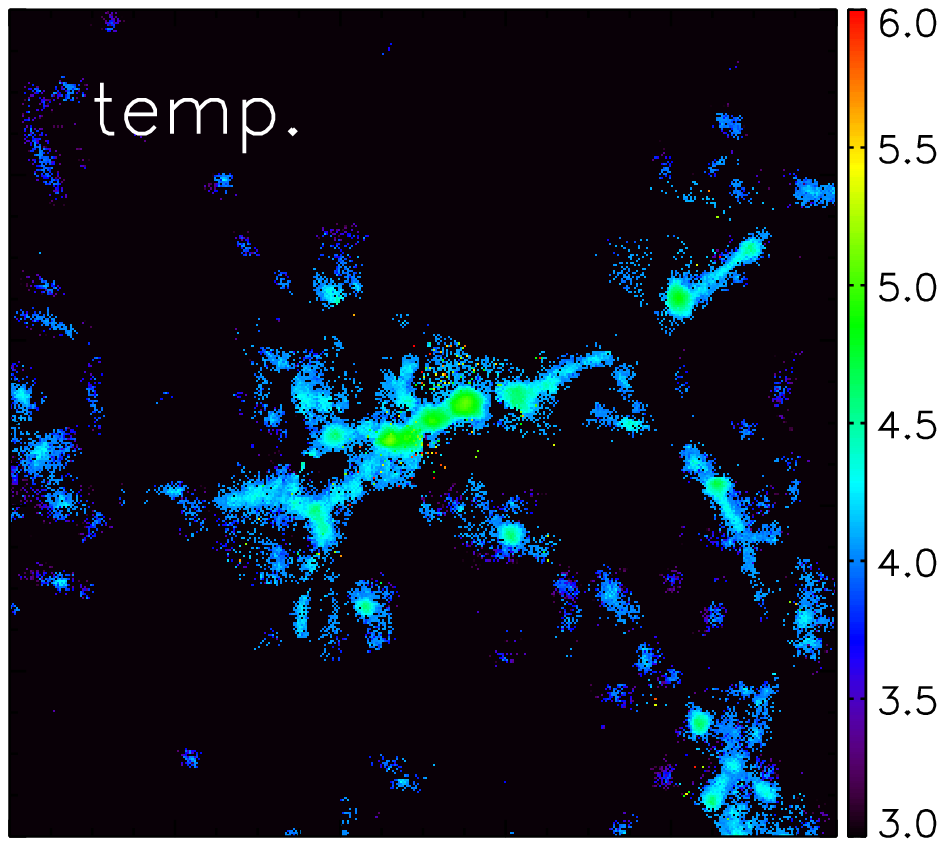}
\includegraphics[scale=0.37, bb=105 165 510 570, clip=true]{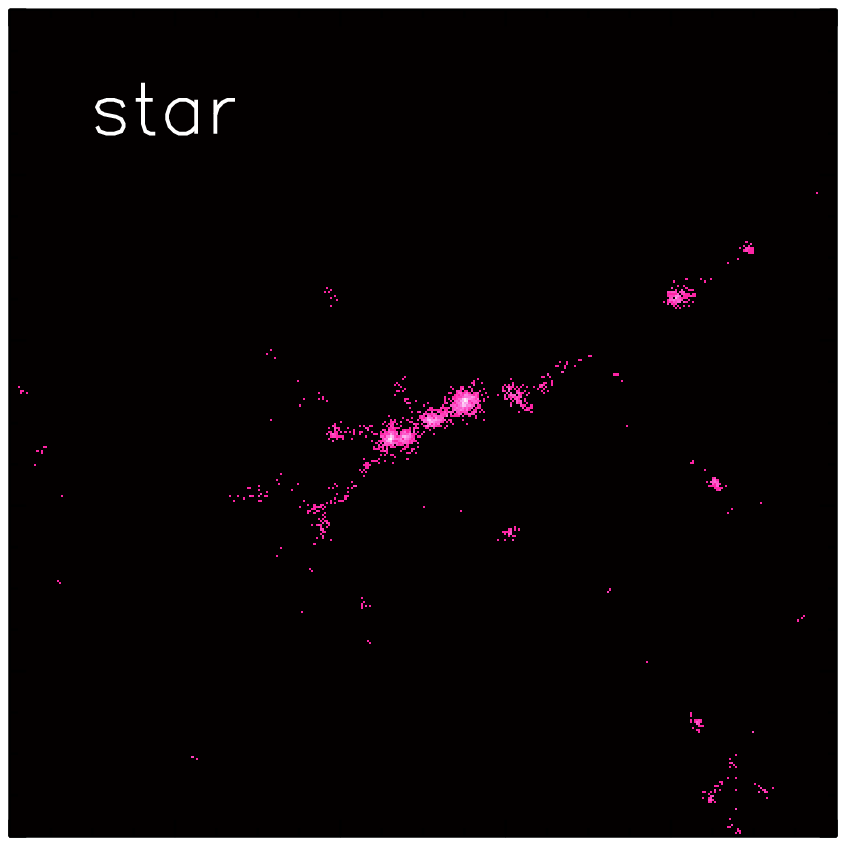}\\
\includegraphics[scale=0.37, bb=105 165 510 570, clip=true]{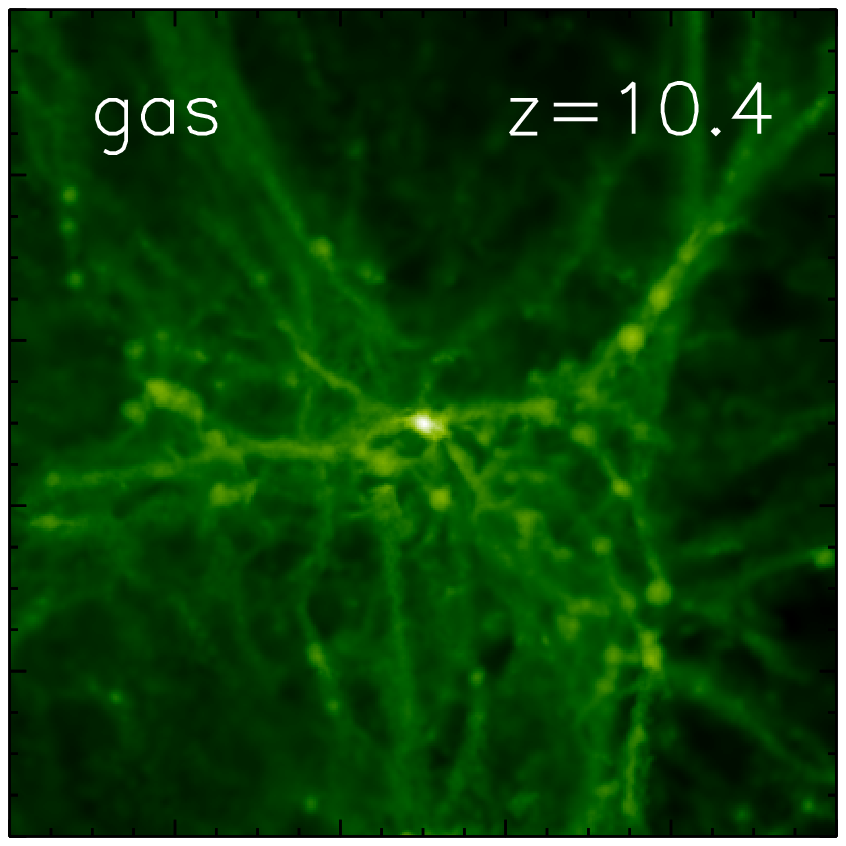}
\includegraphics[scale=0.37, bb=105 165 560 570, clip=true]{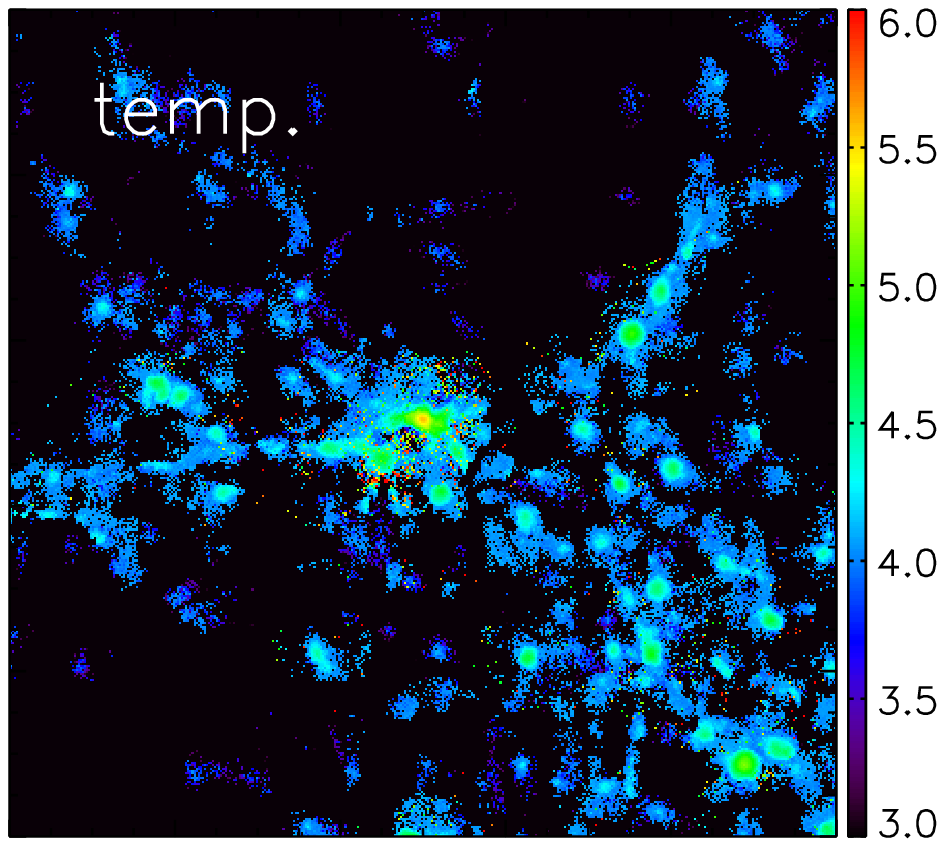}
\includegraphics[scale=0.37, bb=105 165 510 570, clip=true]{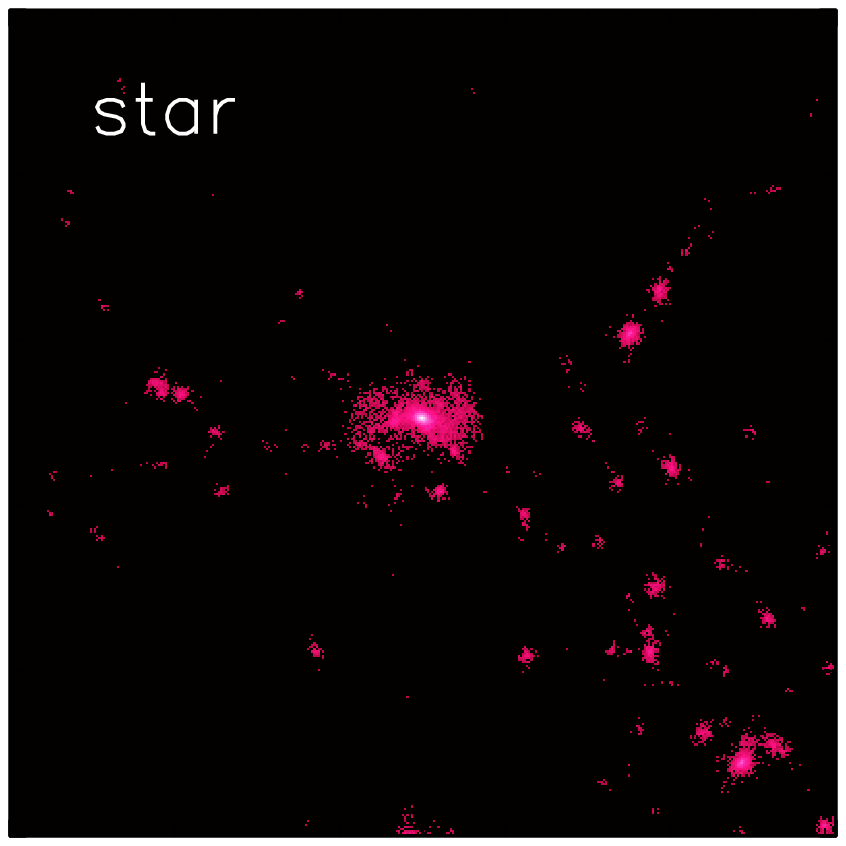}\\
\includegraphics[scale=0.37, bb=105 165 510 570, clip=true]{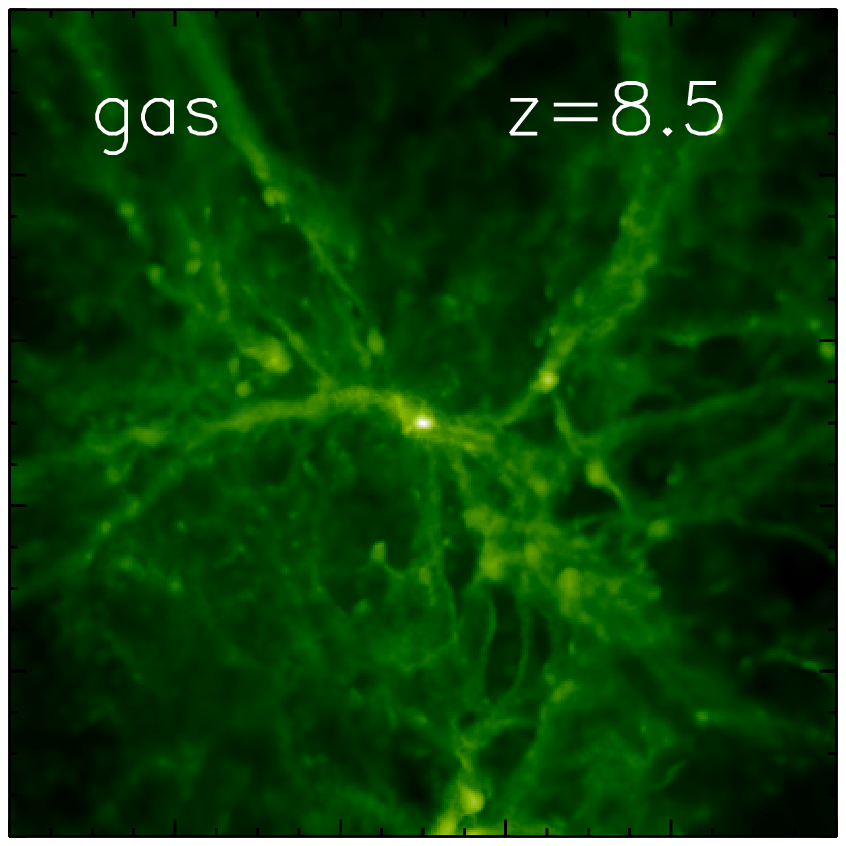}
\includegraphics[scale=0.37, bb=105 165 560 570, clip=true]{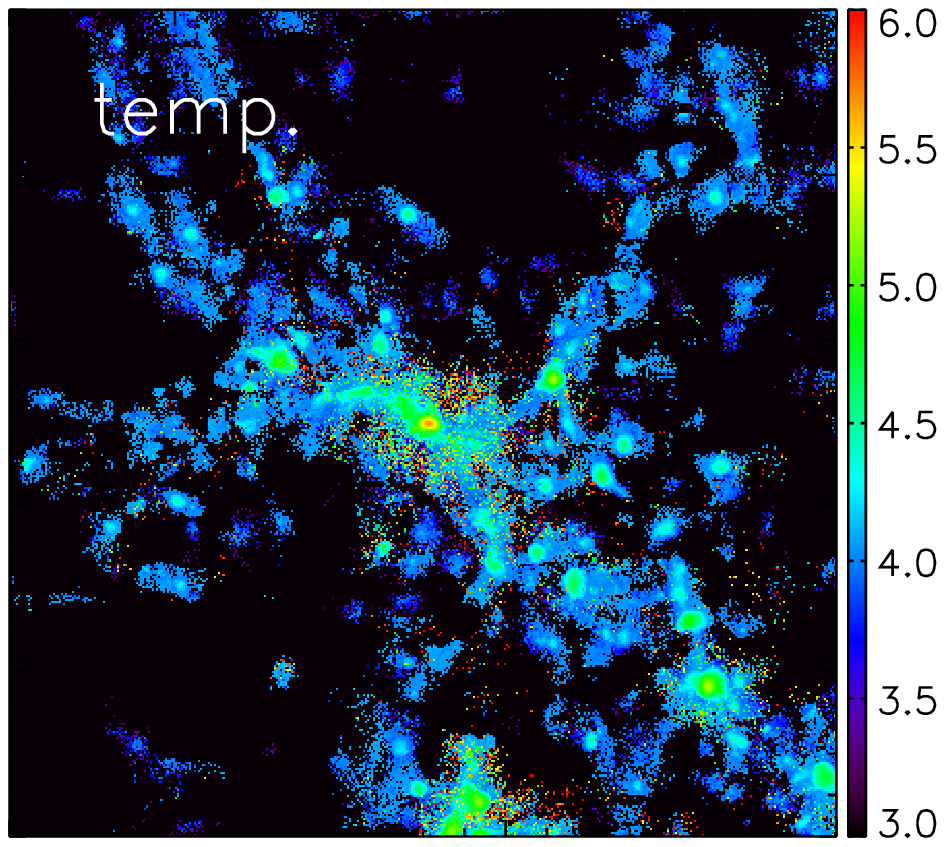}
\includegraphics[scale=0.37, bb=105 165 510 570, clip=true]{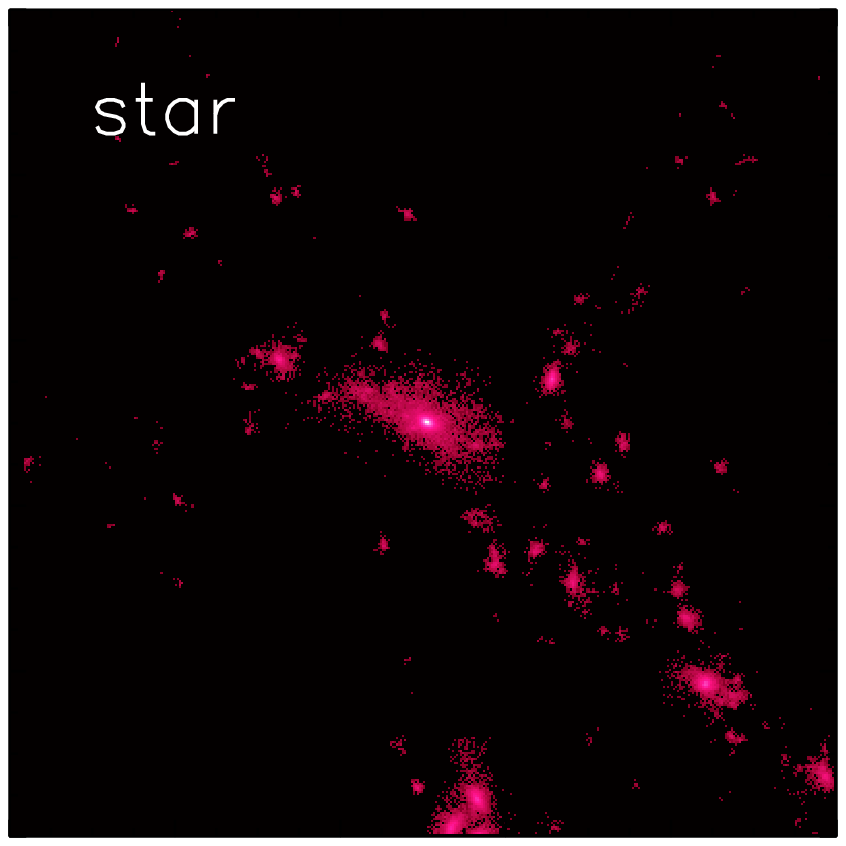}\\
\includegraphics[scale=0.37, bb=105 165 510 570, clip=true]{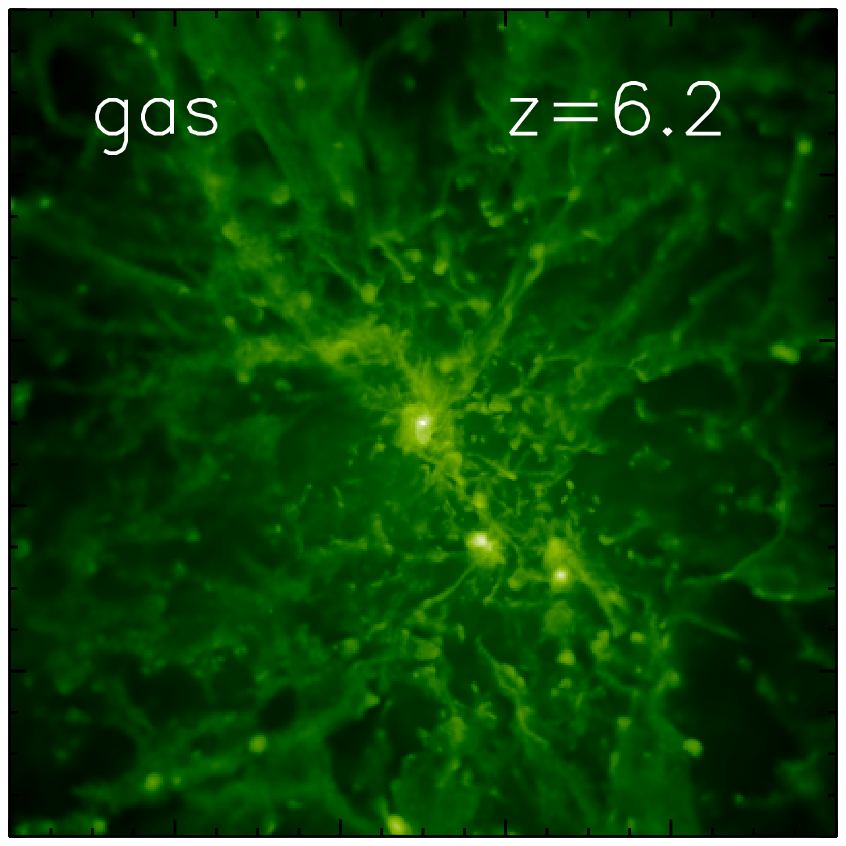}
\includegraphics[scale=0.37, bb=105 165 560 570, clip=true]{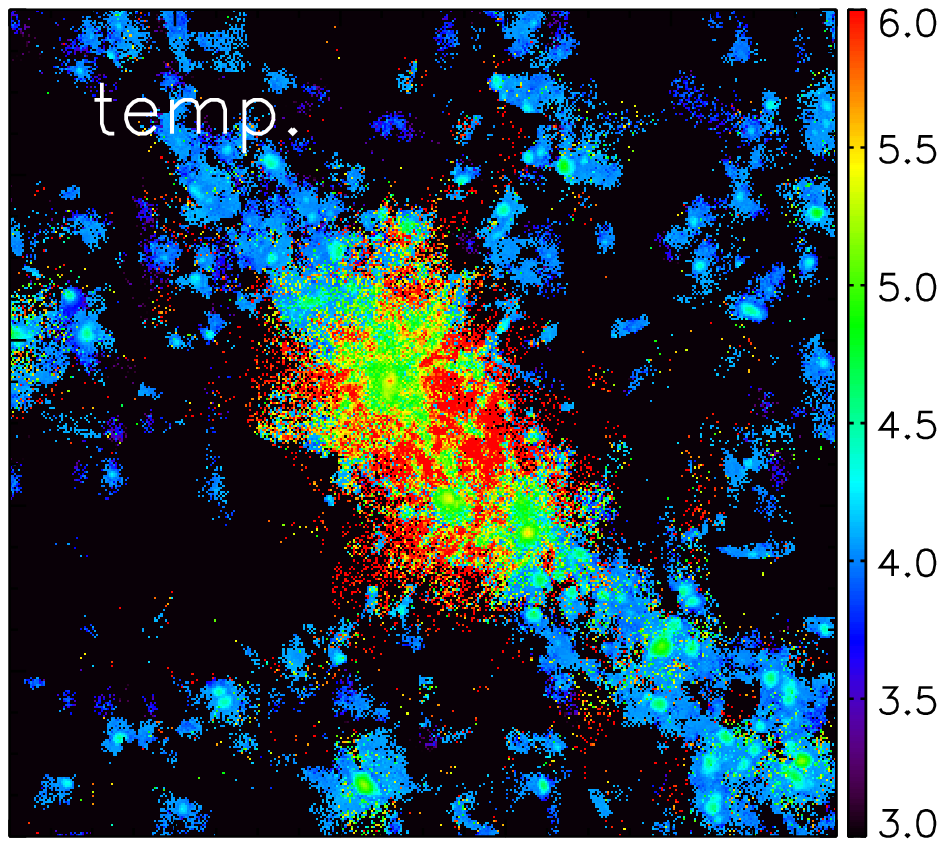}
\includegraphics[scale=0.37, bb=105 165 510 570, clip=true]{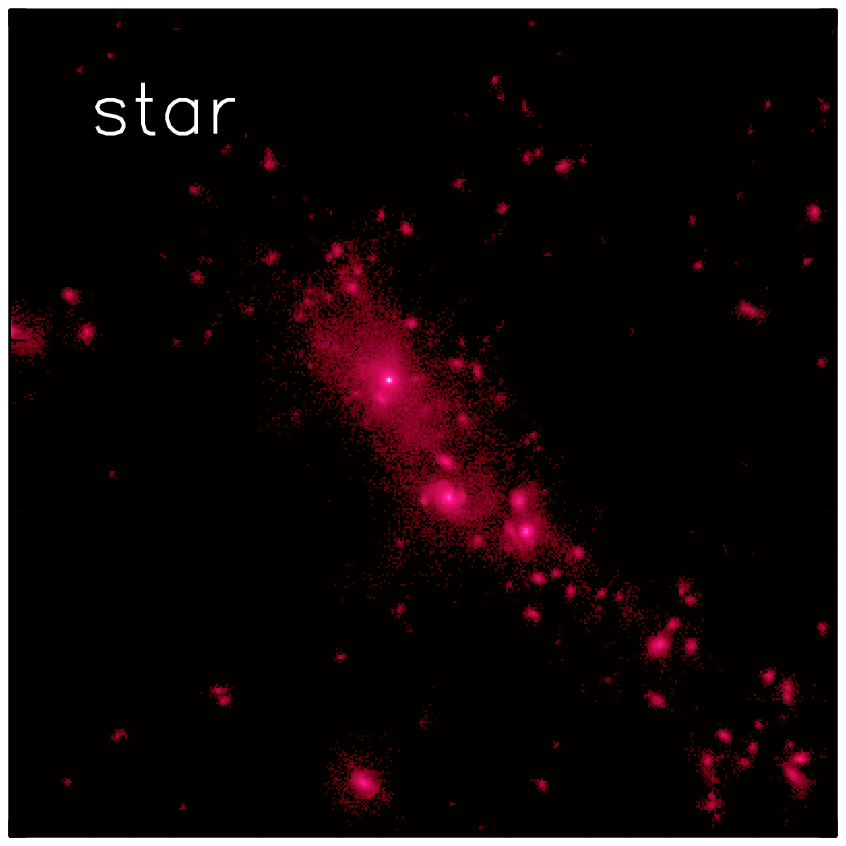}
\caption{The distribution of gas density (left column), gas temperature (middle column),  and stellar density (right column), of the MW galaxy at $z \sim 14, 10.4, 8.5$, and 6.2, respectively. The box size is 1 Mpc in comoving scale. The temperature of the gas is in Kelvin in log scale, as indicated  in the color bar.
}
\label{fig:img}
\end{center}
\end{figure*}

\begin{figure}
\begin{center}
\includegraphics[scale=0.42]{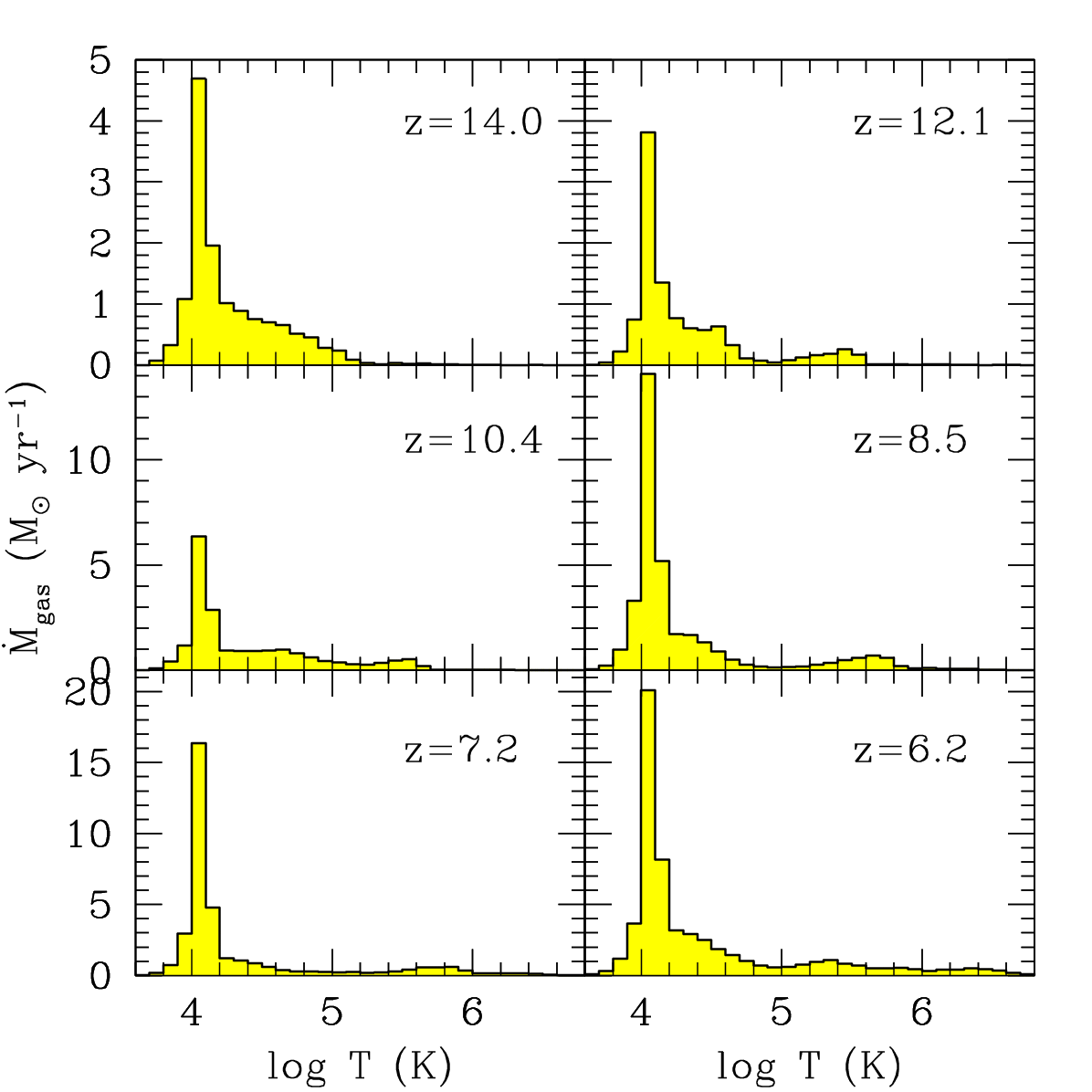}
\caption{The gas accretion rate by the MW galaxy as a function of gas temperature at different redshift.
}
\label{fig:ar}
\end{center}
\end{figure}

\begin{figure}
\begin{center}
\includegraphics[scale=0.42]{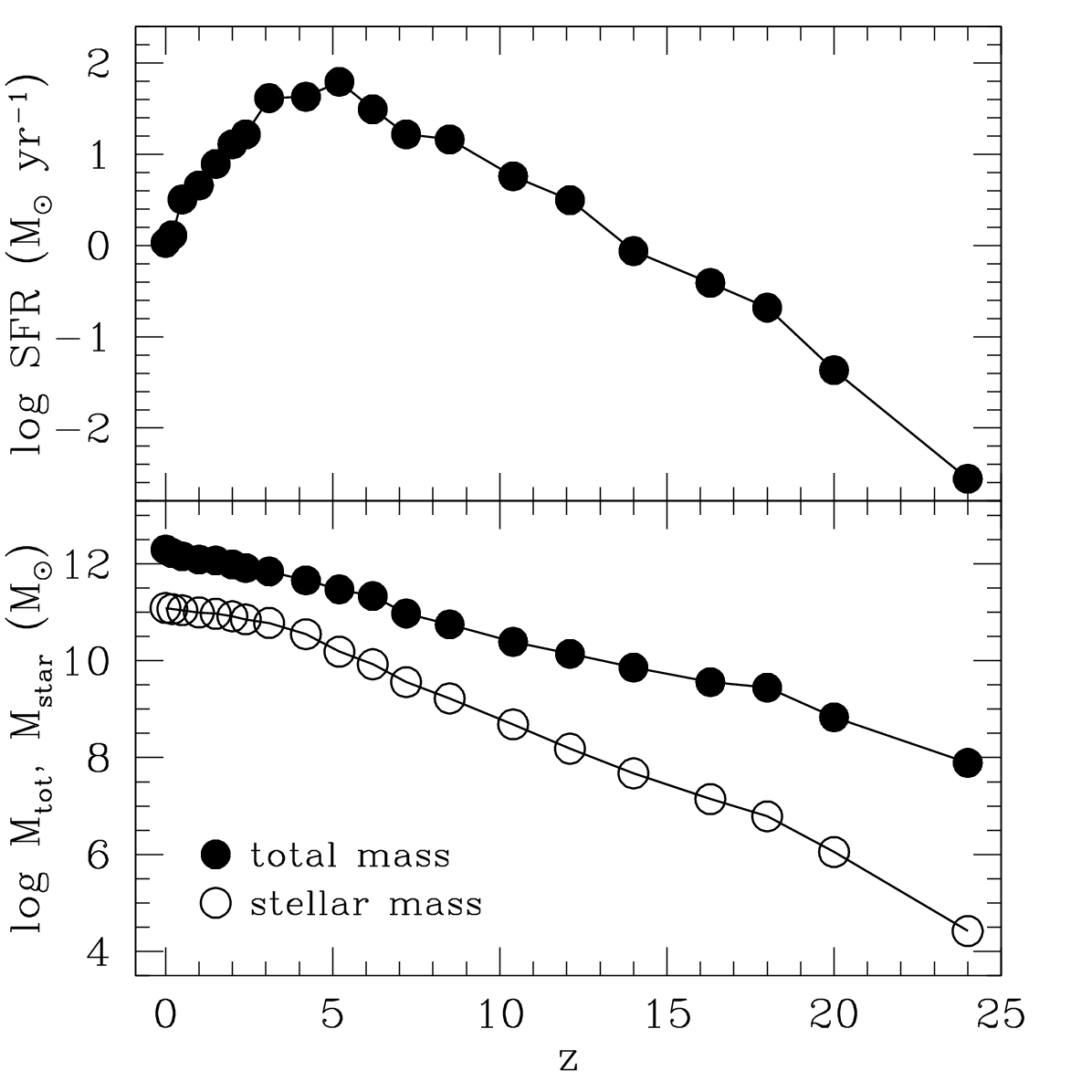}
\caption{The growth history of the MW galaxy illustrated by the star formation rate (top panel), and by the accumulated mass (bottom panel), in which filled circle represents the total mass, while the open circle represents the stellar mass. 
}
\label{fig:sfr}
\end{center}
\end{figure}

The modeled MW galaxy starts to form at $z \sim 24$. Figure~\ref{fig:img} shows the distribution of gas density, gas temperature, and stellar density of the MW main progenitor from redshift $z \sim 14$ to $z \sim 6$. The gas follows the distribution of dark matter and exhibits filamentary structures. At $z \gtrsim 6$, the gas is predominantly cold, with a mean temperature of $\sim 10^4$~K. Stars form out of such cold gas, so they also distribute along the filaments. 

The star formation at $z \gtrsim 6$ is fueled by efficient accretion of cold gas, as demonstrated in Figure~\ref{fig:ar}. The gas accretion rate is defined as the inflow rate of gas within the virial radius of the modeled galaxy. It peaks around $10^4$~K in all cases. At a later time,  feedback from both stars and accreting BHs heats up the gas.
Also, the gas can be heated by gravitational shocks during the infall.
Therefore the accretion includes hot gas as well. The inflow gas falls along the filaments toward the intersection, the highest density peak where the first galaxy in the simulated volume forms. 

Figure~\ref{fig:sfr} shows the star formation history of the MW. The star formation rate (SFR) increases steadily from $\sim 3 \times 10^{-3}~\Msunyr$ at $z \sim 24$ to $\sim 15 ~\Msunyr$ at $z \sim 8.5$, and it peaks at $\sim 62 ~\Msunyr$ at $z = 5.2$, 
due to the merging processes of gas-rich galaxies.
The galaxy mass increases rapidly during this cold accretion phase. By $z \sim 8.5$, it reaches a total mass of $\sim 5.6\times 10^{10}\, \Msun$, and a stellar mass of $\sim 6 \times 10^{9}\, \Msun$.

\subsection{The $\lya$ Properties}

\begin{figure*}
\begin{center}
\includegraphics[scale=0.5, bb=75 135 550 550, clip=true]{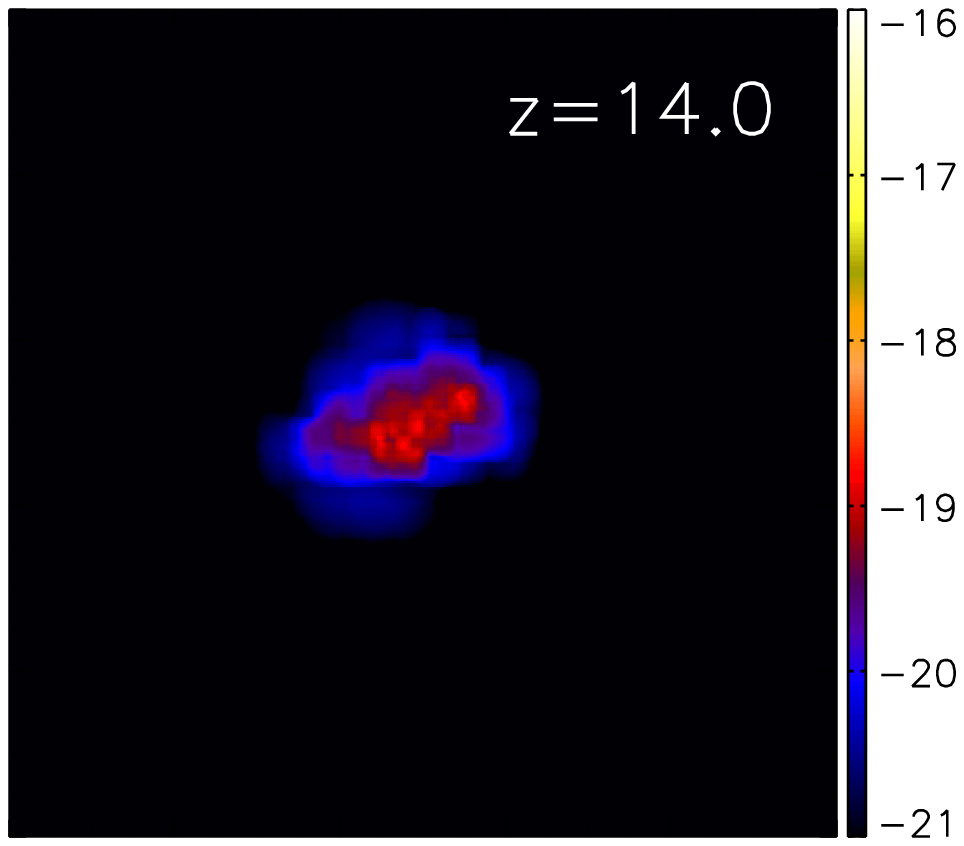}
\includegraphics[scale=0.5, bb=75 135 550 550, clip=true]{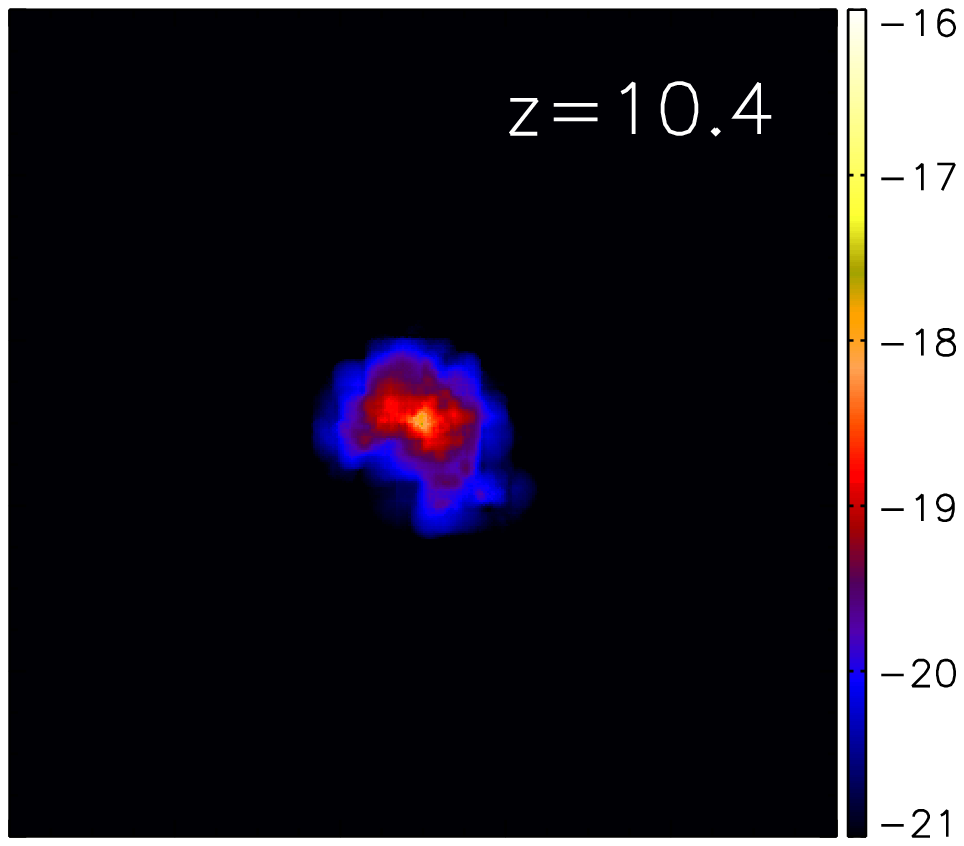}\\
\includegraphics[scale=0.5, bb=75 135 550 550, clip=true]{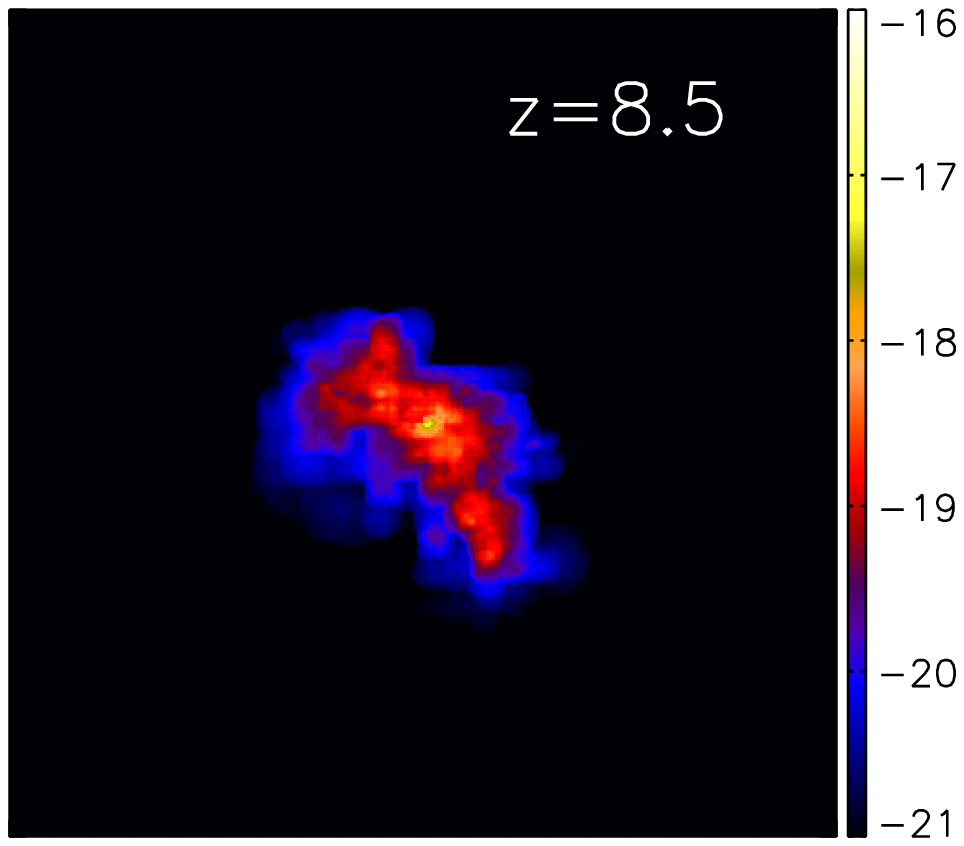}
\includegraphics[scale=0.5, bb=75 135 550 550, clip=true]{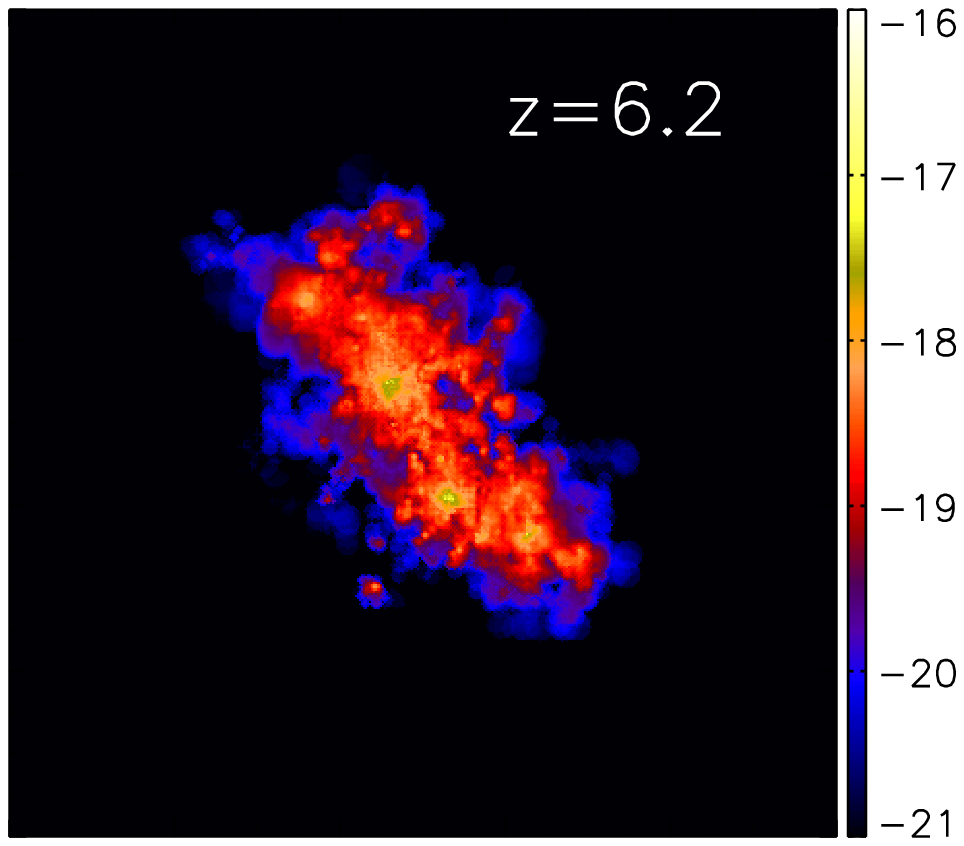}
\caption{The evolution of the $\lya$ surface brightness of the MW galaxy with redshift, at $z \sim 14, 10.4, 8.5$, and 6.2, respectively. The box size is 1 Mpc in comoving scale. The color indicates the $\lya$ surface brightness in log scale in units of $\rm erg~s^{-1}~cm^{-2}~arcsec^{-2}$.
}
\label{fig:sb}
\end{center}
\end{figure*}

\begin{figure}
\begin{center}
\includegraphics[scale=0.45]{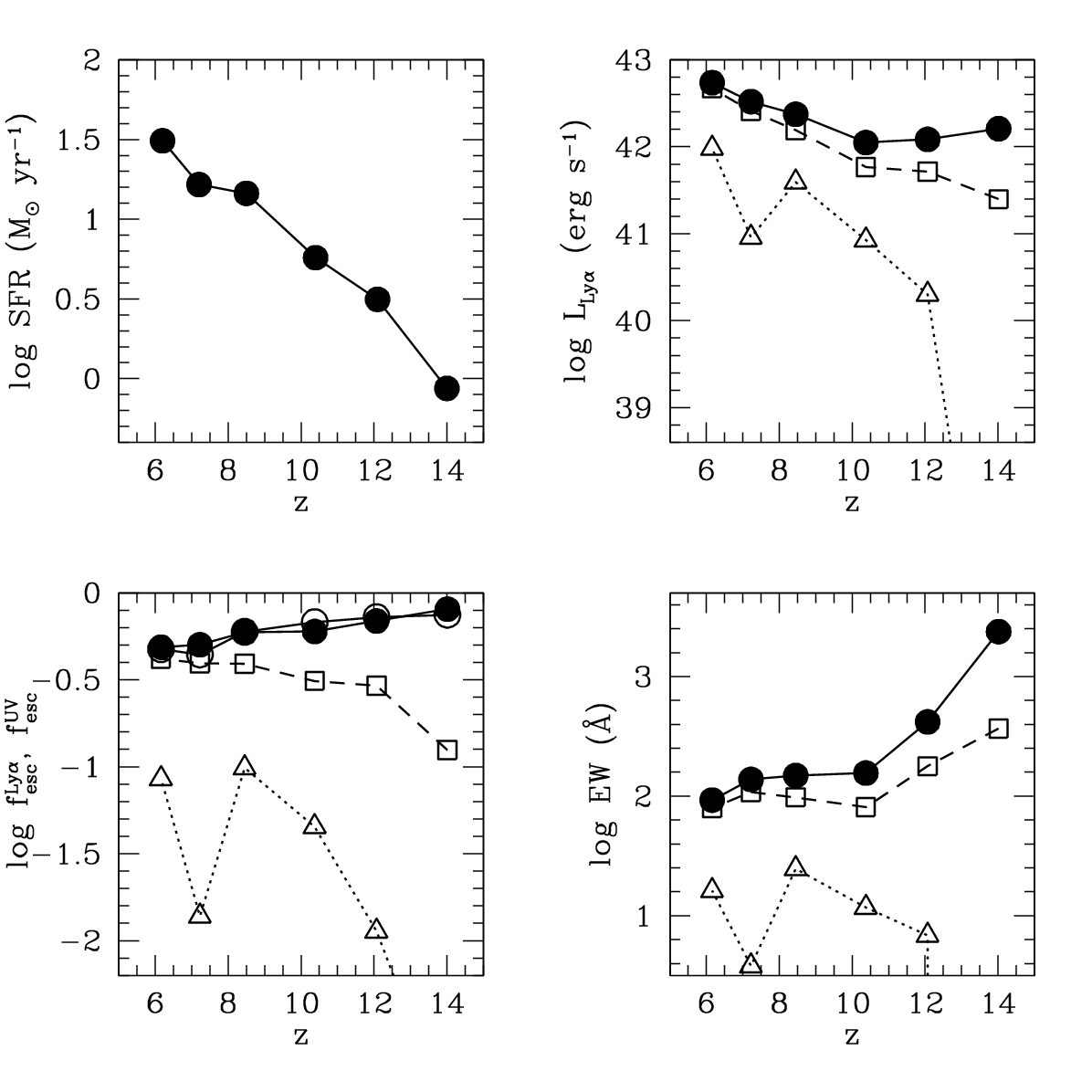}
\caption{
The $\lya$ properties of the modeled galaxy from $z \sim 14$ to $z \sim 6$, including, in clockwise direction, star formation rate, emergent $\lya$ luminosity, equivalent width of $\lya$ line in rest frame, and photon escape fraction of $\lya$ (filled circles) and UV continuum (1300 - 1600 $\A$, open circles).
Open triangles and squares represent the modified $\lya$ properties considering the detection thresholds of the surface brightness with $10^{-18}$ and $10^{-19} ~\rm erg~s^{-1}~cm^{-2}~arcsec^{-2}$, respectively.
}
\label{fig:La}
\end{center}
\end{figure}

The $\lya$ emission traces the gas distribution, as shown in Figure~\ref{fig:sb}. The surface brightness rises above $10^{-20}~\rm{erg~s^{-1}~cm^{-2}~arcsec^{-2}}$ at $z \lesssim 14$. At high redshift $z \gtrsim 10$, the galaxy is small, and the $\lya$ emission is faint and confined to the central high-density region. 
The $\lya$ emission increases with the mass and size of the galaxy, and
  it becomes 
stronger and more extended and irregular due to mergers and gas infall along the filaments of the main halo. 

Figure~\ref{fig:La} shows the $\lya$ properties of the MW galaxy at from $z \sim 14$ to $z \sim 6$, including the emergent $\lya$ luminosity ($\La$), equivalent width of $\lya$ line in rest frame, and photon escape fraction of $\lya$ and UV continuum (1300 - 1600 $\A$). For comparison with the star formation activity, the SFR of the galaxy at corresponding redshift is also shown. 

During this early growth phase, the SFR of the galaxy increases from $\sim 1 ~\Msunyr$ at $z \sim 14.0$ to $\sim 31 ~\Msunyr$ at $z \sim 6$, owing to abundant supply of cold gas from infall and merging of gas-rich mini halos.  The resulting emergent $\lya$ luminosity shows a similar trend, increasing from {$\sim {1.6} \times 10^{42}\, \ergs$} at $z \sim 14.0$ to {$\sim 5.5 \times 10^{42}\, \ergs$}
  at $z \sim 6$. If we consider only the recombination process with the assumption of $\La / L_{\rm H_{\alpha}} = 8.7$ (in which the ${\rm H_{\alpha}}$ is a tracer of star formation), the the intrinsic $\lya$ luminosity should be linearly proportional to SFR, $\La\; (\ergs) = 1.1\times10^{42} \times SFR\; (\Msunyr)$ \citep{Kennicutt98}. However, the evolution of $\La$ in Figure~\ref{fig:La} differs from the SFR history. This is due to the contribution from excitation cooling to the $\lya$ emission as we will discuss later, and dust absorption of the $\lya$ photons. In particular, at $z \gtrsim 10$, the $\La$ increases with redshift, in opposite direction from the SFR, as a result of high collisional excitation and high $\fesc$.

The lower-left panel of Figure~\ref{fig:La} shows the photon escape fraction of $\lya$, $f_{\rm esc}^{\lya}$, and the UV continuum $f_{\rm esc}^{\rm UV}$, where $f_{\rm esc}^{\rm UV}$ is calculated at $\lambda_{\rm rest} = 1300 - {1600} \;\A$. The $\fescalpha$ of the modeled galaxy falls in the range of $0.49 - 0.81$, and it increases with redshift. In our model, the dust is produced by type-II supernovae \citep{Li08}. The dust amount increases as star formation rises from $z \sim14$ to $z \sim 6$, and hence it efficiently absorbs the $\lya$ and UV continuum photons, resulting decreasing escape fraction. However even at $z = 8.5$, about $40\%$ of $\lya$ photons are absorbed by dust.
This is due to the fact that, in the early phase, galaxies are gas rich and compact, the gas and dust are highly concentrated in the galaxies, resulting in effective absorption of the the $\lya$ and UV photons by the dust. 

The resulting $\lya$ equivalent width (EW) is shown in the lower-right panel of Figure~\ref{fig:La}. 
The EW is estimated from the $\lya$ flux divided by the UV flux density at $\lambda = 1300 - {1600} \; \A$.
The modeled galaxy has $\EW \gtrsim 20\; \A$ at these redshifts, and is therefore classified as a $\lya$ emitter (LAE) \citep[e.g.,][]{Gronwall07}. The EW increases with redshift, from {$\sim 93~\A$} at $z \sim 6$ to {$\sim 2300~\A$} at $z \sim 14.0$.
Such trend is similar to that reported in recent observations which showed that galaxies at
higher redshift have higher EWs than their lower-redshift counterparts \citep[e.g.,][]{Gronwall07, Ouchi08}.
This is because the contribution from excitation $\lya$ cooling becomes large with increasing redshift, as shown in Figure~\ref{fig:exc} in the next section, which boosts the EW significantly \citep{Yajima12b, Yajima12c}. 

The currently most distant LAE at $z=7.5$, z8\_GND\_5296, has a $\lya$ luminosity of $\sim 1.8 \times 10^{42}\, \rm{erg\, s^{-1}}$ \citep{Finkelstein13}.
Our model shows the similar $\lya$ luminosity at the redshift, and hence may be reproducing the observed LAE. 
However, there are additional uncertainties which may reduce the $\lya$ flux of our calculation as we explain at the below and Section~\ref{sec:profile}. 

Next generation telescopes will have very high-angular resolution,
for example,  that of JWST will get to $\lesssim 0.1"$.
Some extended fainter parts as seen in Figure~\ref{fig:sb} can be lost in observation with such the resolution.
As a result, observed $\lya$ flux is likely to be lower than that galaxies are actually emitting. 
In practice, for Figure~\ref{fig:sb}, if we count up only fluxes of pixels brighter than $10^{-18}~\rm erg \; s^{-1} \; cm^{-2} \; arcsec^{-2}$, 
which is the detection threshold of recent observation of extended $\lya$ source with a narrow-band filter \citep[e.g.,][]{Matsuda12}, 
the $\lya$ fluxes are reduced by factor $\sim 5.7$ at $z=6.2$ and $\sim 61.0$ at $z=12.1$. 
Open triangles and squares in the Figure~\ref{fig:La} show the $\lya$ properties by considering the detection thresholds of the surface brightness with $10^{-18}$ and $10^{-19} ~\rm erg~s^{-1}~cm^{-2}~arcsec^{-2}$, respectively.
$\lya$ luminosity and $\EW$ of our model galaxies can be significantly reduced in the mock observation with the threshold of $10^{-18} ~\rm erg~s^{-1}~cm^{-2}~arcsec^{-2}$. 
In particular, the galaxies at $z > 10$ become too faint to be detected in the current observation. 
On the other hand, if the surface brightness threshold is $\sim 10^{-19}~\rm erg \; s^{-1} \; cm^{-2} \; arcsec^{-2}$, 
we detect $85.9$ per cent of the flux for $z=6.2$ and $42.5$ per cent for $z=12.1$.
However, the narrow-band filter imaging by F164N on JWST will require very long exposure time $\gtrsim 10^{4}$ hours to achieve the detection threshold for $5\; \sigma$ detection. 
Therefore, although galaxies have complex $\lya$ distribution reflecting gas and stellar distribution, 
most of them can be lost in observation. 
Due to the lost of faint extended parts, the high-$z$ galaxies can be faint at $\lya$ band or undetectable with current (or future) observations, 
although they are intrinsically bright. 

\subsection{Contribution of Excitation Cooling to $\lya$ Emission}
\label{sec:exc}

\begin{figure}
\begin{center}
\includegraphics[scale=0.4]{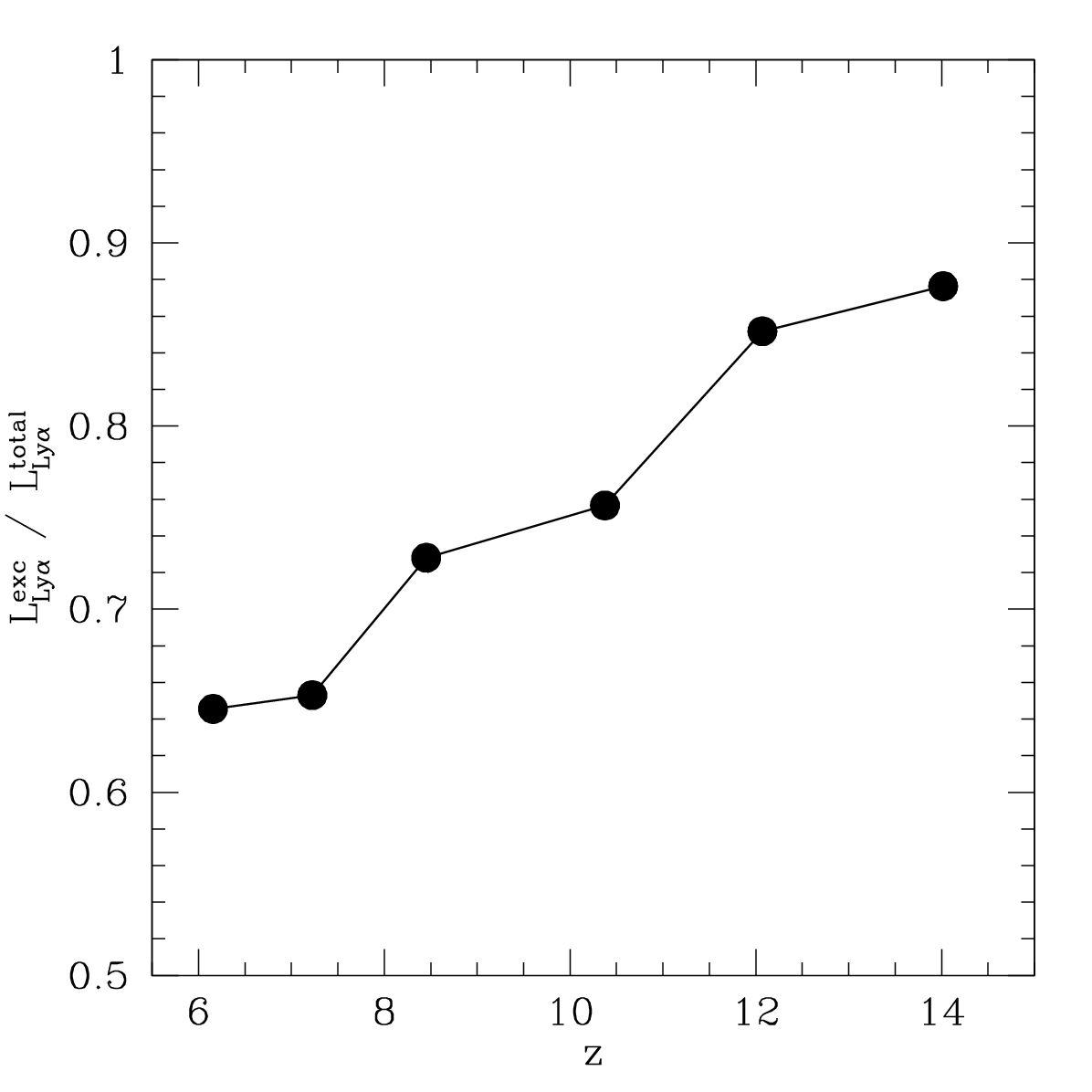}
\caption{
The fraction of excitation cooling $\lya$ to the total intrinsic $\lya$ luminosity as a function of redshift from $z \sim 14$ to $z \sim 6$.
}
\label{fig:exc}
\end{center}
\end{figure}

As mentioned in \S~\ref{sec:rt}, the $\lya$ emission is generally produced by the recombination of ionizing photons and the collisional excitation of hydrogen gas. In our cosmological simulation, galaxy evolution is accompanied by cold, filamentary gas streams with temperature $T\sim 10^{4-5}~\rm K$, which penetrate deep inside the dark matter halos \citep[Zhu et al. in prep,][]{Yajima12c}, 
which was also shown by previous theoretical works
\citep{Katz03, Keres05, Keres09, Birnboim03, Dekel06, Ocvirk08, Brooks09, Dekel09}. 
The $\lya$ emissivity due to collisional excitation is sensitive to gas temperature, and has the peak in the efficiency at $T\sim 10^{4}~\rm K$ \citep{Faucher09}. 
Hence, much excitation $\lya$ cooling photons can be emitted from such cold accreted gas \citep{Dijkstra09, Faucher09, Goerdt10}. 
At higher redshift, galaxies experience more merging events and accrete more cold gas efficiently, which results in stronger $\lya$ emission from excitation cooling, and higher $\lya$ EWs \citep{Yajima12b, Yajima12c}.

As shown in Figure~\ref{fig:exc}, the fraction of intrinsic excitation cooling $\lya$ to the total intrinsic $\lya$ luminosity increases from {$\sim 65\%$} at $z \sim 6$ to {$\sim 88\%$} at $z \sim 14$. Such extremely high  excitation $\lya$ cooling produces the extremely high $\lya$ EWs seen in Figure~\ref{fig:La}. 

The $\lya$ luminosity of our model, which is mainly contributed by the excitation cooling, is higher than the model at $z=3$ in \citet{Faucher10}. 
For example, they showed $\La \lesssim 10^{42}~\ergs$ at the halo mass $M_{\rm h} \sim 10^{11}~\Msun$. 
On the other hand, when our model galaxy has similar mass at $z = 7.2$, it shows $\La = 6.6 \times 10^{42}~\ergs$ without dust extinction. 
This may be due to the difference of the conversion efficiency from gravitational energy to $\lya$ cooling. 
\citet{Faucher10} used 0.3 as the conversion efficiency \citep[see also][]{Dijkstra09}. 
In addition, recently \citet{Rosdahl12} showed the conversion efficiency is $\sim 0.1-0.2$ by radiative-hydrodynamics simulations. 
However, this conversion efficiency depends sensitively on the detailed gas structure in and around galaxies \citep{Rosdahl12}.
In some situations, cold-accreted gas is disturbed by interstellar gas and heated up \citep{Rosdahl12}.  
In our model, galaxies are compact and high-density \citep{Yajima12d}, hence a large fraction of accreted gas might be heated due to friction with interstellar medium.
Then, since the temperature of the cold accretion gas is $\gtrsim 10^{4}~\rm K$, most of the thermal energy can be converted to $\lya$ photons \citep{Thoul96}.
In addition, $\lya$ luminosity by the excitation cooling can increases with redshift, because the $\lya$ emissivity is proportional to square of gas density (Equation 1) and mean gas density of galaxies increases with redshift \citep[e.g.,][]{Bryan98}.
In practice, \citet{Goerdt10} showed that $\La = 1.88 \times 10^{42}~\ergs~(M_{\rm h} / 10^{12}~\Msun)^{0.8} (1+z)^{1.3}$ by their cosmological hydrodynamics simulations with a simple dust absorption model. 
The $\lya$ luminosities of our model galaxies are consistent with their estimation. 
Note that, the $\lya$ cooling rate balances heating rate and it is sensitive to temperature. 
The heating rate may not simply increase with redshift while the mean gas density does. 
If gas temperature is higher than $\sim 10^{6}~\rm K$, the thermal energy can be released by different cooling radiation, 
e.g., recombination, free-free emission \citep{Thoul96}. 
In addition, the $\lya$ luminosity by the excitation cooling in our simulations at $z=3$, which is $\La=1.9\times10^{42}~\ergs$ at $M_{\rm h}=5.9\times 10^{11}~\Msun$, 
is close to that in \citet{Faucher10}. 
Thus, the $\lya$ luminosity of our model at $z > 6$ can be higher than the analytical model of \citet{Faucher10} at $z = 3$ by some factors. 

The current our code does not distinguish excitation and recombination $\lya$ photons in the RT calculations. 
However, the fraction of excitation $\lya$ cooling rate may not change significantly for mock observations with the different thresholds of surface brightness. 
This is because, as shown in \citet{Yajima12c}, $\lya$ photons are mostly emitted at galactic centers, and travel with many scatterings in inter-stellar medium, resulting in the faint extended parts. 
Therefore, the mock observation with the different thresholds of surface brightness can miss the both excitation and recombination $\lya$ photons at the faint parts. 

\subsection{The $\lya$ Line Profile}
\label{sec:profile}

\begin{figure}
\begin{center}
\includegraphics[scale=0.44]{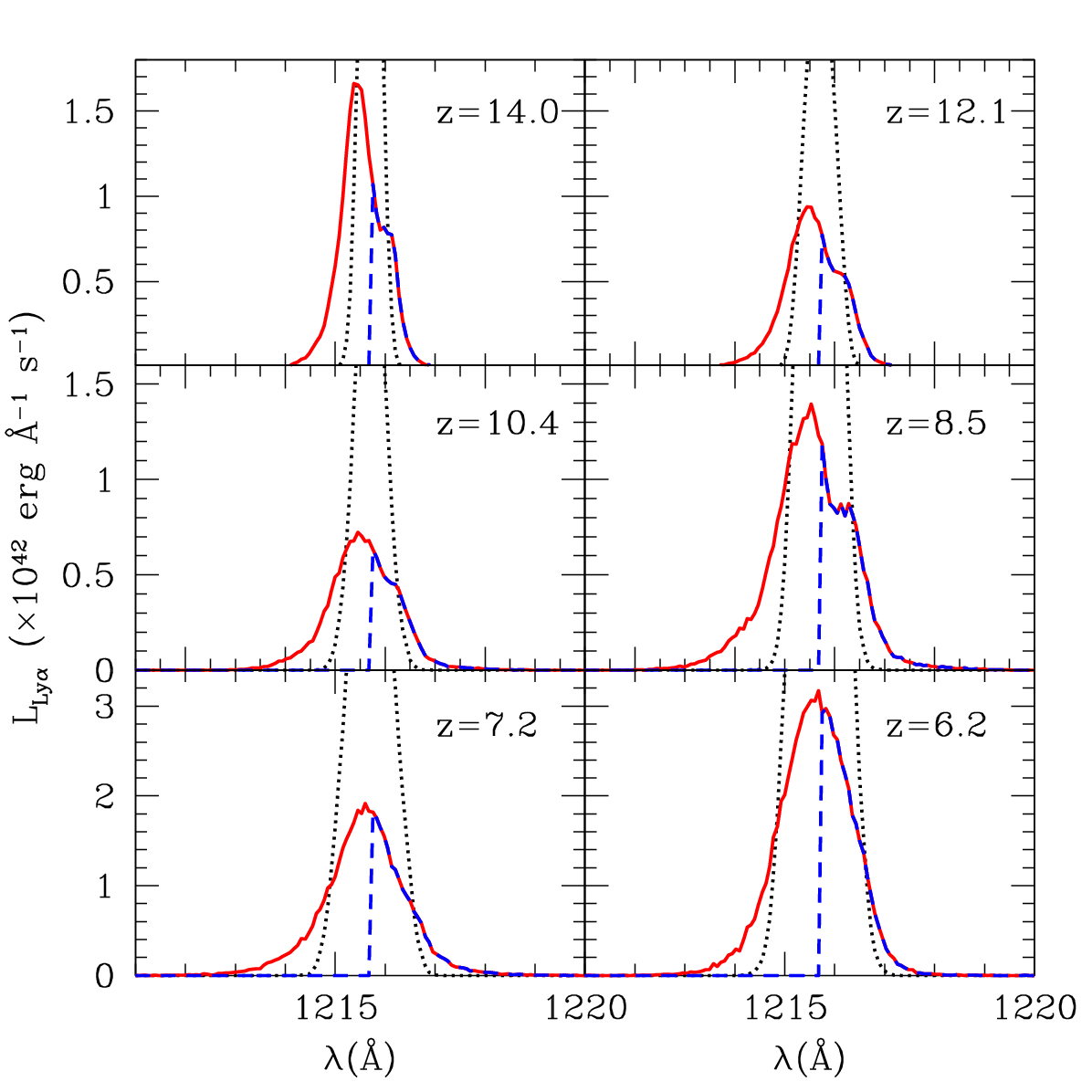}
\caption{
The $\lya$ line profile of the modeled galaxy at different redshifts.
The black dotted- and  red solid lines are the intrinsic and emergent $\lya$ profiles, respectively.
The blue dash lines are the $\lya$ profiles with cutting off $\lya$ flux at $\lambda < \rm 1216 ~\AA$
due to IGM transmission \citep{Laursen11}.
}
\label{fig:profile}
\end{center}
\end{figure}

\begin{figure}
\begin{center}
\includegraphics[scale=0.44]{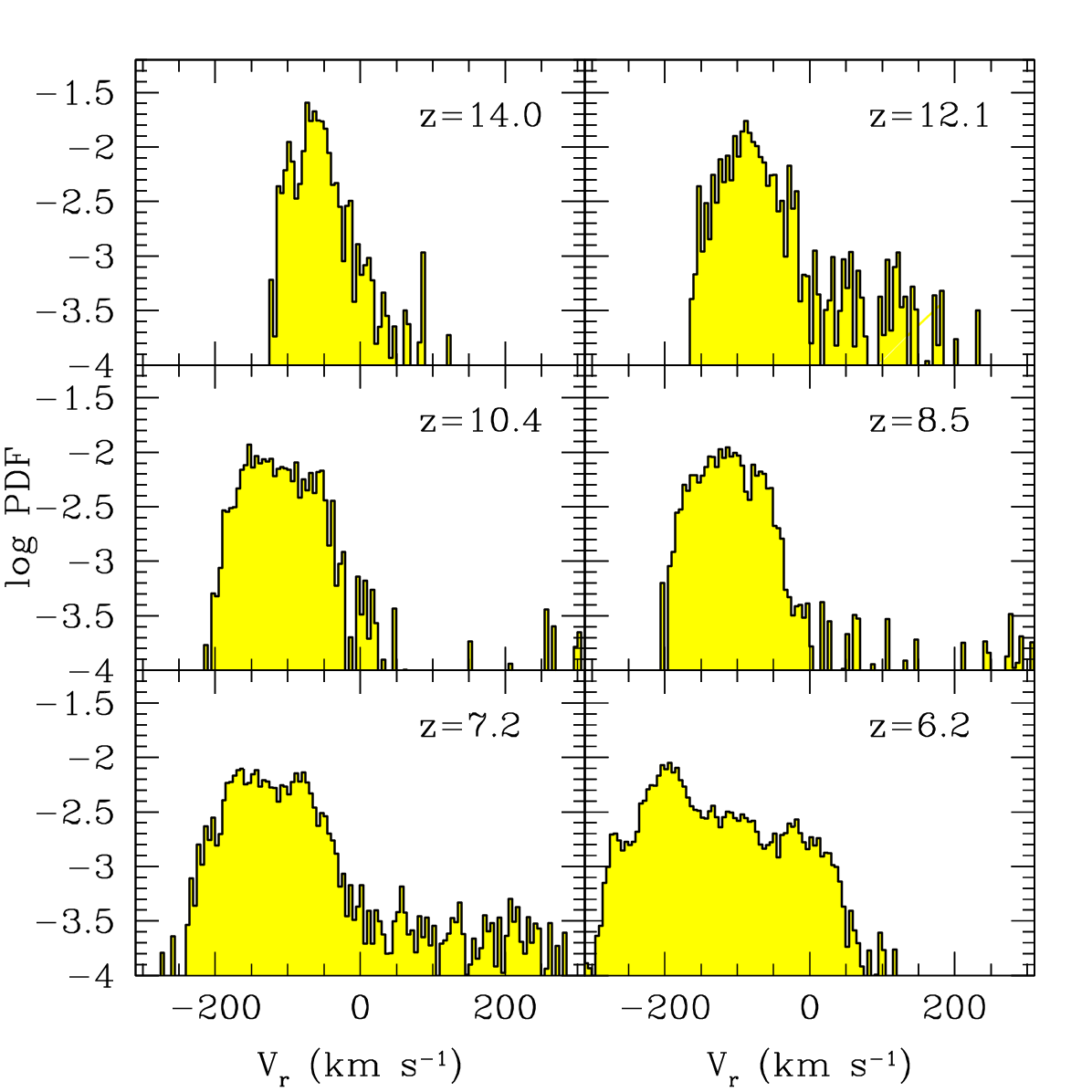}
\caption{
The probability distribution function of the gas mass of the neutral hydrogen in the galaxy as a function of radial velocity. The velocity is estimated from the center of mass of the galaxy in the radial direction.
}
\label{fig:vmap}
\end{center}
\end{figure}

The resulting $\lya$ line profiles of the modeled MW galaxy from $z \sim 14$ to $z \sim 6$ are shown in Figure~\ref{fig:profile}. 
We randomly sample the frequency of the intrinsic $\lya$ photon from a Maxwellian distribution with the gas temperature at the emission location. 
All sources show asymmetric profiles with a {single} peak or weak double peaks. More interestingly, most profiles are shifted to the shorter (bluer) wavelengths. This is a characteristic feature of gas inflow \citep{Zheng02}. Indeed, as shown in Figure~\ref{fig:vmap}, a significant fraction of the gas shows a large infalling velocity $V_{\rm r} \sim -100$ to $\sim - 200~\rm{km~s^{-1}}$, even though our simulation includes feedback of stellar wind similar to that of \citet{Springel05d}. In particular, the gas in the galaxy from $z \sim 14 - 10$ is dominated by inflow motion, which explains the significant blue shift of the profiles in Figure~\ref{fig:profile} (top panel). At redshift $z \lesssim 8.5$, the gas exhibits outflow as well, and has a larger velocity distribution $ - 250 \lesssim V_{\rm r} \lesssim 200~\rm km~s^{-1}$, which results in an extended profile to both blue and red wings. 
While asymmetric line profiles with an extended red wing are commonly seen in high-redshift LAEs, there appears to be some profiles in the $z \gtrsim 6$ observations that have complex features including double peaks and extended blue wing, similar to what we see here \citep[e.g.,][]{Ouchi10, Hu2010, Kashikawa2011}. 
The observed line of z8\_GND\_5296, the most distant LAE at $z=7.5$, is not resolved well and thus has a Gaussian profile \citep{Finkelstein13}. 
More observations of high-resolution $\lya$ line profiles of high-redshift LAEs are needed to test our model and verify our predictions. 

We note that the $\lya$ line profile may be suppressed and changed by the intergalactic medium (IGM) \citep[e.g.,][]{Santos04, Dijkstra07, Zheng10, Laursen11}, because the IGM effectively scatters the $\lya$ photons at the line center and at shorter wavelengths by the Hubble flow \citep[e.g.,][]{Laursen11}. 
As a result, the inflow feature in our profiles may disappear and the shape may become an asymmetric single peak with only photons at red wing. 
\citet{Laursen11} showed that a large fraction of $\lya$ flux from galaxies at $z \sim 6.5$ could be lost by scattering in IGM
despite most of IGM were ionized.
As a simple test, we show the line profiles without photons at the shorter wavelength as shown in the blue dash lines in the figure. 
As a result, about 0.59 (0.54) of $\lya$ flux from the galaxies at $z=12.1$ (6.2) are lost. 
The inflow feature completely disappears, and the asymmetric profiles with the red wings may be recognized as the galaxies have gas outflow.
In addition, if IGM is highly neutral, even $\lya$ flux at red wing is highly suppressed. 
For neutral IGM, the IGM optical depth is estimated by $\tau (\Delta v) \sim 2.3 \left( \frac{\Delta v}{600\rm~km/s}\right) \left( \frac{1+z}{10} \right)^{3/2}$ 
\citep{Dijkstra10}, where $\Delta v$ is the velocity shift from the line center.
More than 0.99 of $\lya$ fluxes from our model galaxies are lost for the neutral IGM. 
Therefore, if IGM is highly neutral, $\lya$ flux from our model galaxies cannot be observed.

\subsection{Detectability of Progenitors of  Local $L^*$ Galaxies}

\begin{figure}
\begin{center}
\includegraphics[scale=0.45]{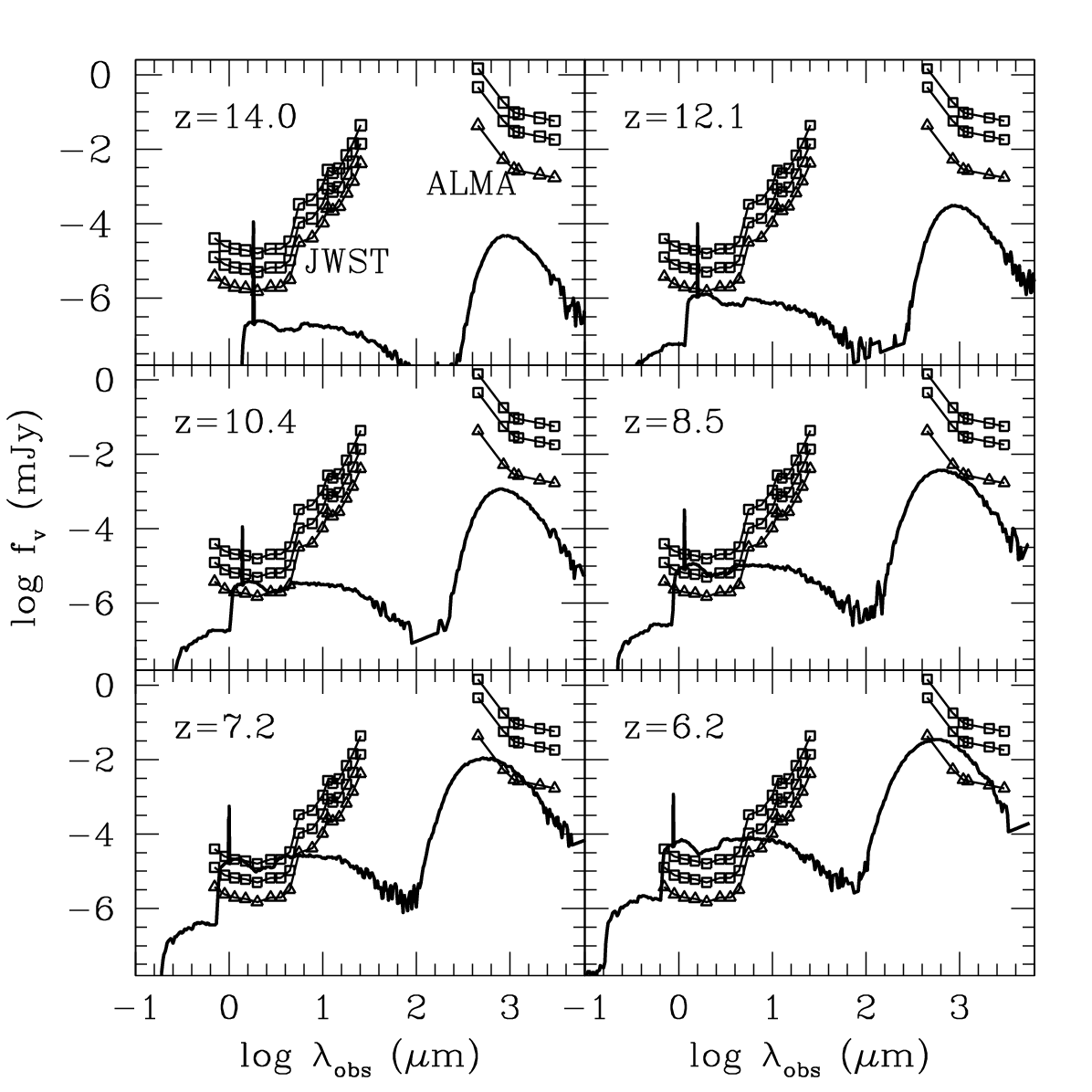}
\caption{
The spectral energy distribution of the MW galaxy at different redshift and its detectability with JWST and ALMA. The open squares indicate the 10~$\sigma$ detection limits of JWST and ALMA with 16 antennas at an integration time of 1 and 10 hours (from top to bottom), while the open triangles indicate the 3~$\sigma$ detection limits of JWST and ALMA with 50 antennas. 
}
\label{fig:sed}
\end{center}
\end{figure}

The emergent multi-wavelength SEDs of
the MW galaxy at different redshifts are shown in Figure~\ref{fig:sed}. 
The shape of the SEDs evolves with redshift due to the change of intrinsic stellar radiation, absorption of continuum photons by gas and dust, and thermal emission by dust. 
In all cases, the strong $\lya$ lines emerge, and at $z \gtrsim 8.5$ the UV continuum at $\lambda \le 912~\A$ in rest frame is deeply declined due to strong absorption because of dense neutral hydrogen gas around star forming region. 

A major science goal of the two forthcoming telescopes, ALMA and JWST, is to detect the first galaxies. In order to predict the detectability of the infancy of a local $L^*$ galaxy, we contrast the SEDs with some detection limits of these two facilities in Figure~\ref{fig:sed}. Our calculations show that the flux at $850~\rm \mu m$ in the observed frame of the model galaxy ranges from $\sim 7.9\times10^{-5}$ mJy at $z = 14.0$ to $\sim 4.7\times 10^{-2}$ mJy at $z=6.2$. With an array of 50 antennas and an integration of 10 hours, ALMA may be able to detect such galaxies at $z \lesssim 8.5$ with a 3~$\sigma$ significance. However, since galaxies do not have a lot of young stars and much dust at $z \gtrsim 10$, observations in continuum by ALMA becomes more difficult, and it would need tens of hours of integration time. In contrast, JWST appears to be more powerful to detect the earliest galaxies as the one we model here, because it can detect the UV continuum in rest frame up to {$z \sim 10$}. 
The $\lya$ emission is strong even at {$z \sim 12$}, which may be observable by Near-Infrared Spectrograph (NIRSpec) on JWST. 
The NIRSpec will have the detection threshold of $\sim 3 \times 10^{-18}~\rm erg \; s^{-1} \; cm^{-2}$ with $R = 100$ and $\rm S/N=10$ by exposure time of $10^{4}~$ seconds.

We note that in the above estimation, IGM absorption and transmission were not taken into account. The IGM can significantly suppress the $\lya$ flux, and the transmission highly depends on viewing angle \citep[e.g.,][]{Laursen11} by inhomogeneous ionization structure in IGM \citep[e.g.,][]{Abel07, Yoshida07, Jeeson-Daniel12}, which make the detection more difficult. Of course, the galaxies we present here represent progenitor of a local $L^*$ galaxy such as the Milky Way. Galaxies formed in highly overdense regions are likely much more massive \citep{Li07}, and may be more easily detected by both ALMA and JWST (Li et al, in preparation).
 
\section{Discussions}

\begin{figure}
\begin{center}
\includegraphics[scale=0.44]{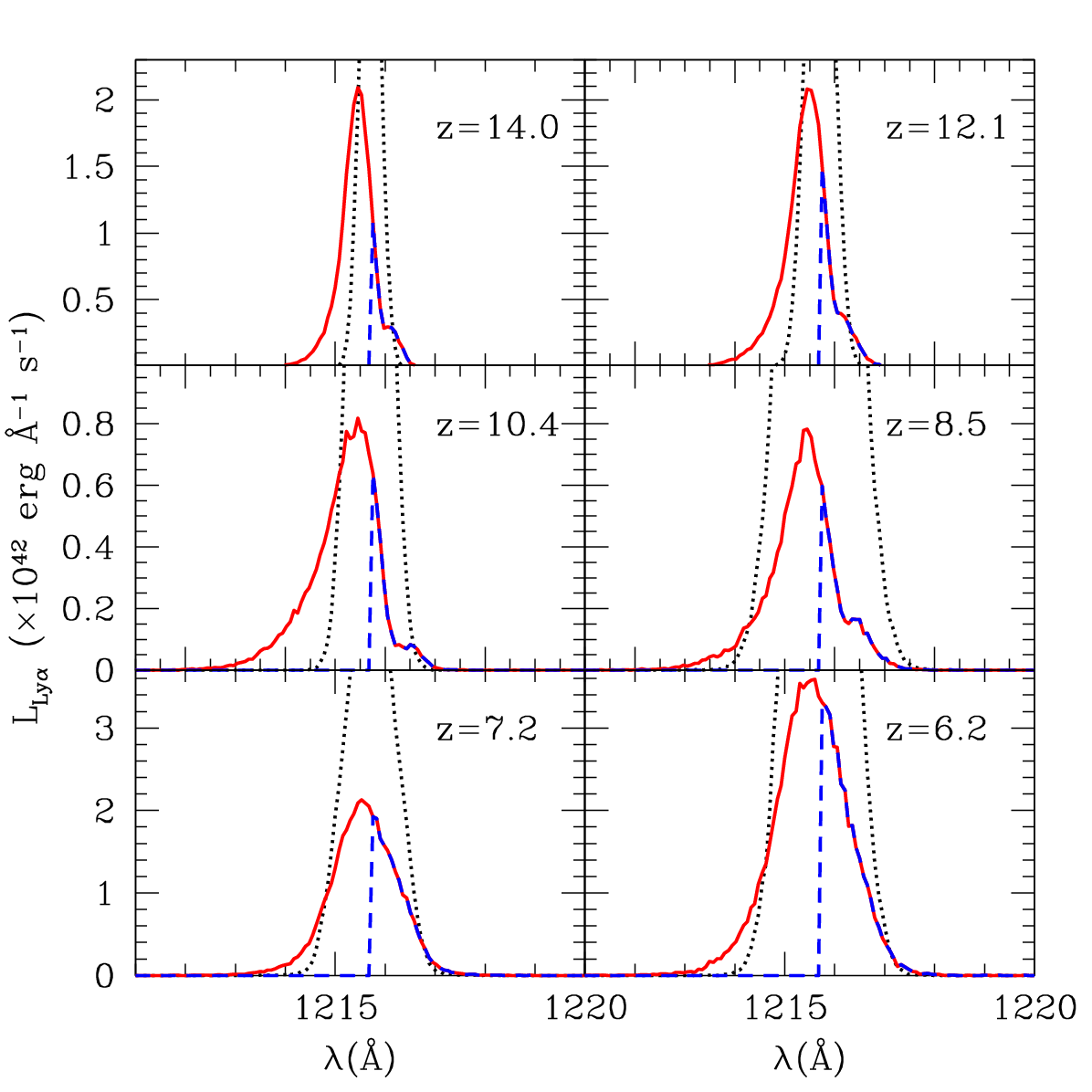}
\caption{Same as in Figure~\ref{fig:profile}, but here the simulation does not include the wind model from stellar feedback.
}
\label{fig:nowindprofile}
\end{center}
\end{figure}

\begin{figure}
\begin{center}
\includegraphics[scale=0.44]{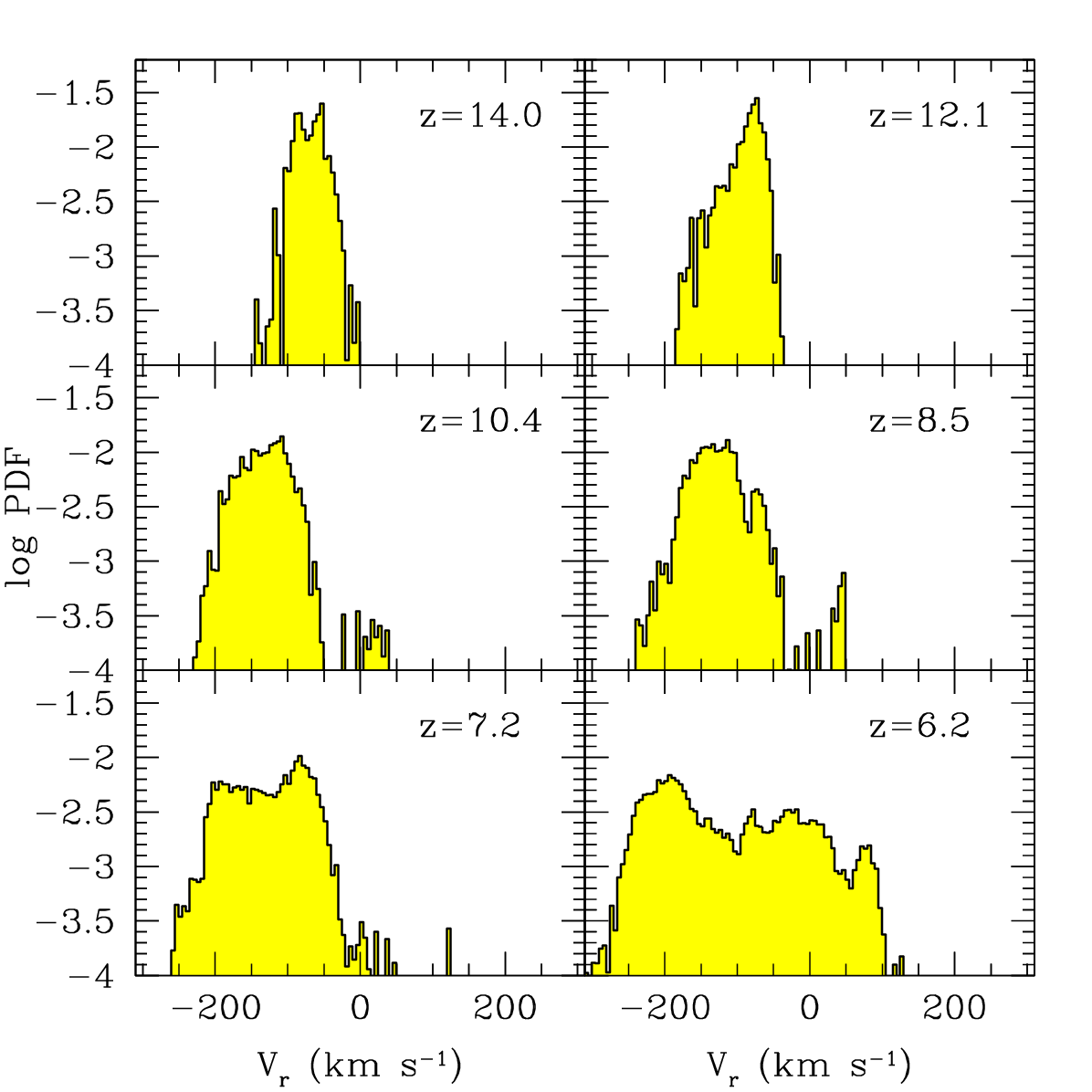}
\caption{
Same as Figure~\ref{fig:vmap}, but here the simulation does not include the wind model from stellar feedback. The probability distribution function of the gas mass of the neutral hydrogen in the galaxy as a function of radial velocity. The velocity is estimated from the center of mass of the galaxy in the radial direction.
}
\label{fig:nowindvmap}
\end{center}
\end{figure}

The gas inflow feature is present in our simulation with outflow from stellar feedback. In order to probe the effect of stellar wind on the gas inflow, we also performed the simulation with a pure thermal feedback model, in which the feedback from supernovae is only in thermal energy. In such a model, some fraction of thermal energy can quickly escape as cooling radiation before conversion to kinetic energy. As a result, gas outflow does not occur efficiently. 

The resulting $\lya$ line profiles and the probability distribution function of the neutral gas mass are shown in Figure~\ref{fig:nowindprofile} and Figure~\ref{fig:nowindvmap}, respectively. Without a strong outflow, the $\lya$ line profiles show more pronounced blue wing.  
However, due to the IGM transmission, $\lya$ flux at the blue wing can be suppressed significantly. 
If $\lya$ photons at shorter wavelength than the line center are suppressed, 
0.69 (0.44) of $\lya$ flux from the galaxies at $z=12.1~ (6.3)$ is lost (see the blue dash lines in Figure~\ref{fig:nowindprofile}). 

The galaxy in our simulation resides in a low overdensity region and it represents those that would evolve into present-day $L^*$ galaxies such as the MW, so it may not be the very first one formed in the universe. It is believed that the most massive halos in the highly overdense regions collapse first, in which the first stars form \citep[e.g.,][]{Abel2002, Bromm2004, Gao2007, Yoshida2008}. These may also be the formation sites of the very first galaxies, owing to feedback and chemical enrichment from the PopIII stars, as well as abundant gas supply \citep{Li07, Bromm2009}. 

One of the major limitations of our model is that the cosmological simulation does not have sufficient resolutions to follow the formation and evolution of individual stars. Instead, star formation is modeled using a ``sub-grid'' recipe based on the observed Schmidt-Kennicutt Law \citep{Kennicutt98}. Gas particles are converted into stars once it is cooled below $10^4$~K and the density is above a threshold \citep{Springel03a}. Although this treatment is rather simplistic, it nevertheless gives a global star formation history close to what is believed of the MW galaxy. 

Another major limitation is that our current RT calculations do not include the propagation and scattering of $\lya$ and ionizing photons in the IGM. We make the prediction that galaxies with inflow of cold gas would result in asymmetric, blue-shifted $\lya$ line profiles. However, as discussed earlier, the absorption by IGM may change the the profile to one with extended red wing. We will study this issue in more detail in future work that includes the radiative transfer of $\lya$ and ionizing photons in the IGM.

%
%

\section{Summary}

In this work, we have investigated the formation of a typical, nearby $L^*$ galaxy such as the MW, and its $\lya$ properties at the earliest evolutionary stage. We combine a cosmological hydrodynamic simulation, which uses the Aquila initial condition and focuses on a MW-like galaxy, with three-dimensional radiative transfer calculations using the improved $\art$ code, which couples multi-wavelength continuum, $\lya$ line, and ionization of hydrogen. 

We find that the modeled MW galaxy forms from efficient accretion of cold gas early on, which sustains a high star formation rate from $z \sim 14 - 6$. The cold accretion produces strong $\lya$ emission via collisional excitation, which has a luminosity from {$\sim {1.6} \times 10^{42}\, \ergs$} at $z \sim 14$ to {$\sim 5.5 \times 10^{42}\, \ergs$} at $z \sim 6$. The escape fraction of $\lya$ photons increases from $\sim 0.49$ at $z \sim 6$  to $\sim 0.81$ at $z \sim 14$, due to less dust content at higher redshift. The EWs of the $\lya$ lines increases with redshift, from {$\sim 93~\A$} at $z \sim 6$ to {$\sim 2300~\A$} at $z \sim 14$. Such high EWs may be due to significant contribution to $\lya$ emission by excitation cooling, which dominates at high redshift. The resulting $\lya$ lines exhibit asymmetric, mostly single-peak profiles shifted to the blue wing, a characteristic feature of inflow. 

Furthermore, we demonstrate that progenitors of local $L^*$ galaxies such as the modeled one may be detected at $z \lesssim 8$ by JWST and ALMA with a reasonable integration time. At higher redshift $z \gtrsim 12$, however, only $\lya$ line may be observable by spectroscopic surveys with similar detection limit as JWST. 

Our results suggest that $\lya$ line may be used to probe the formation and evolution, and gas properties of distant galaxies. It is perhaps one of the most powerful tools to detect the first generation of galaxies in the coming decade.

%
%
\acknowledgments

We thank Carlos Frenk for kindly providing the Aquila initial conditions to
us, and Masayuki Umemura for stimulating discussions and helpful
comments.
We also thank the anonymous referee for useful comments.
HY is grateful to Takatoshi Shibuya, Alex Hagan, and Robin Ciardullo for 
valuable discussion about the sensitivity of JWST. 
This research is supported by NSF grants AST-0965694, AST-1009867, and
AST-0807075. We thank the Research Computing and Cyberinfrastructure 
(unit of Information Technology Services) at the Pennsylvania State University for
providing computational resources and services. The Institute for Gravitation
and the Cosmos (IGC) is supported by the Office of the Senior Vice President for
Research and the Eberly College of Science at the Pennsylvania State University.  

%
%


\begin{thebibliography}{74}
\expandafter\ifx\csname natexlab\endcsname\relax\def\natexlab#1{#1}\fi

\bibitem[{{Abel} {et~al.}(2002){Abel}, {Bryan}, \& {Norman}}]{Abel2002}
{Abel}, T., {Bryan}, G.~L., \& {Norman}, M.~L. 2002, Science, 295, 93

\bibitem[{{Abel} {et~al.}(2007){Abel}, {Wise}, \& {Bryan}}]{Abel07}
{Abel}, T., {Wise}, J.~H., \& {Bryan}, G.~L. 2007, \apjl, 659, L87

\bibitem[{{Beichman} {et~al.}(2012)}]{Beichman12}
{Beichman}, C.~A. and {Rieke}, M. and {Eisenstein}, D., et al. 2012, Proceedings of the SPIE, Space Telescopes and Instrumentation 2012: Optical, Infrared, and Millimeter Wave., Volume 8442

\bibitem[{{Birnboim} \& {Dekel}(2003)}]{Birnboim03}
{Birnboim}, Y. \& {Dekel}, A. 2003, 
  \mnras, 345, 349

\bibitem[{{Bouwens} \& {Illingworth}(2006)}]{Bouwens2006}
{Bouwens}, R.~J. \& {Illingworth}, G.~D. 2006,  \nat, 443, 189

\bibitem[{{Bouwens} {et~al.}(2011){Bouwens}, {Illingworth}, {Labbe}, {Oesch},
  {Trenti}, {Carollo}, {van Dokkum}, {Franx}, {Stiavelli}, {Gonz{\'a}lez},
  {Magee}, \& {Bradley}}]{Bouwens2011A}
{Bouwens}, R.~J., {Illingworth}, G.~D., {Labbe}, I., et al. 2011,  \nat, 469, 504

\bibitem[{{Bouwens} {et~al.}(2010){Bouwens}, {Illingworth}, {Oesch},
  {Stiavelli}, {van Dokkum}, {Trenti}, {Magee}, {Labb{\'e}}, {Franx},
  {Carollo}, \& {Gonzalez}}]{Bouwens2010}
{Bouwens}, R.~J., {Illingworth}, G.~D., {Oesch}, P.~A., et al. 2010,  \apjl,
  709, L133

\bibitem[{{Bouwens} {et~al.}(2004){Bouwens}, {Thompson}, {Illingworth},
  {Franx}, {van Dokkum}, {Fan}, {Dickinson}, {Eisenstein}, \&
  {Rieke}}]{Bouwens2004}
{Bouwens}, R.~J., {Thompson}, R.~I., {Illingworth}, G.~D., et al. 2004, \apjl, 616, L79

\bibitem[{{Bromm} \& {Larson}(2004)}]{Bromm2004}
{Bromm}, V. \& {Larson}, R.~B. 2004, \araa, 42, 79

\bibitem[{{Bromm} \& {Yoshida}(2011)}]{Bromm2011}
{Bromm}, V. \& {Yoshida}, N. 2011, \araa, 49, 373

\bibitem[{{Bromm} {et~al.}(2009){Bromm}, {Yoshida}, {Hernquist}, \&
  {McKee}}]{Bromm2009}
{Bromm}, V., {Yoshida}, N., {Hernquist}, L., \& {McKee}, C.~F. 2009, \nat, 459, 49

\bibitem[{{Brooks} {et~al.}(2009){Brooks}, {Governato}, {Quinn}, {Brook}, \&
  {Wadsley}}]{Brooks09}
{Brooks}, A.~M., {Governato}, F., {Quinn}, T., {Brook}, C.~B., \& {Wadsley}, J.
  2009, \apj, 694,
  396

\bibitem[{{Bruzual} \& {Charlot}(2003)}]{Bruzual03}
{Bruzual}, G. \& {Charlot}, S. 2003, \mnras, 344, 1000

\bibitem[{{Bryan} \& {Norman}(1998)}]{Bryan98}
Bryan G. L., Norman M. L., 1998, \aj, 495, 80

\bibitem[{{Dav{\'e}} {et~al.}(1999){Dav{\'e}}, {Hernquist}, {Katz}, \&
  {Weinberg}}]{Dave99}
{Dav{\'e}}, R., {Hernquist}, L., {Katz}, N., \& {Weinberg}, D.~H. 1999, ApJ, 511, 521

\bibitem[{{Dekel} \& {Birnboim}(2006)}]{Dekel06}
{Dekel}, A. \& {Birnboim}, Y. 2006, \mnras, 368, 2

\bibitem[{{Dekel} {et~al.}(2009){Dekel}, {Birnboim}, {Engel}, {Freundlich},
  {Goerdt}, {Mumcuoglu}, {Neistein}, {Pichon}, {Teyssier}, \&
  {Zinger}}]{Dekel09}
{Dekel}, A., {Birnboim}, Y., {Engel}, G., et al. 2009, \nat, 457, 451

\bibitem[{{Di Matteo} {et~al.}(2011){Di Matteo}, {Khandai}, {DeGraf}, {Feng},
  {Croft}, {Lopez}, \& {Springel}}]{DiMatteo11}
{Di Matteo}, T., {Khandai}, N., {DeGraf}, C., et al. 2012, \apjl, 745, L29

\bibitem[{{Dijkstra} {et~al.}(2007){Dijkstra}, {Lidz}, \&
  {Wyithe}}]{Dijkstra07}
{Dijkstra}, M., {Lidz}, A., \& {Wyithe}, J.~S.~B. 2007, \mnras, 377, 1175

\bibitem[{{Dijkstra} \& {Loeb}(2009)}]{Dijkstra09}
{Dijkstra}, M. \& {Loeb}, A. 2009, \mnras, 400, 1109

\bibitem[{{Dijkstra} \& {Wyithe}(2010)}]{Dijkstra10}
{Dijkstra}, M. \& {Wyithe}, J.~S.~B.  2010, \mnras, 408, 352

\bibitem[{{Faucher-Gigu{\`e}re} {et~al.}(2010){Faucher-Gigu{\`e}re}, {Kere{\v
  s}}, {Dijkstra}, {Hernquist}, \& {Zaldarriaga}}]{Faucher10}
{Faucher-Gigu{\`e}re}, C., {Kere{\v s}}, D., {Dijkstra}, M., {Hernquist}, L.,
  \& {Zaldarriaga}, M. 2010, \apj, 725, 633

\bibitem[{{Faucher-Gigu{\`e}re} {et~al.}(2009){Faucher-Gigu{\`e}re}, {Lidz},
  {Zaldarriaga}, \& {Hernquist}}]{Faucher09}
{Faucher-Gigu{\`e}re}, C., {Lidz}, A., {Zaldarriaga}, M., \& {Hernquist}, L.
  2009, \apj, 703, 1416

\bibitem[{{FInkelstein} {et~al.}(2013)}]{Finkelstein13}
{Finkelstein}, S.~L., {Papovich}, C., {Dickinson}, M., et al. 2013, \nat, 502, 524

\bibitem[{{Gao} {et~al.}(2007){Gao}, {Yoshida}, {Abel}, {Frenk}, {Jenkins}, \&
  {Springel}}]{Gao2007}
{Gao}, L., {Yoshida}, N., {Abel}, T., et al. 2007, \mnras, 378, 449

\bibitem[{{Goerdt} {et~al.}(2010){Goerdt}, {Dekel}, {Sternberg}, {Ceverino},
  {Teyssier}, \& {Primack}}]{Goerdt10}
{Goerdt}, T., {Dekel}, A., {Sternberg}, A., et al. 2010, \mnras, 407, 613

\bibitem[{{Greif} {et~al.}(2010){Greif}, {Glover}, {Bromm}, \&
  {Klessen}}]{Greif2010}
{Greif}, T.~H., {Glover}, S.~C.~O., {Bromm}, V., \& {Klessen}, R.~S. 2010, \apj, 716,
  510

\bibitem[{{Gronwall} {et~al.}(2007){Gronwall}, {Ciardullo}, {Hickey},
  {Gawiser}, {Feldmeier}, {van Dokkum}, {Urry}, {Herrera}, {Lehmer}, {Infante},
  {Orsi}, {Marchesini}, {Blanc}, {Francke}, {Lira}, \& {Treister}}]{Gronwall07}
{Gronwall}, C., {Ciardullo}, R., {Hickey}, T., et al.  2007,  \apj, 667, 79

\bibitem[{Haardt \& Madau(1996)}]{Haardt96}
Haardt, F. \& Madau, P. 1996, ApJ, 461, 20

\bibitem[{{Hernquist} \& {Katz}(1989)}]{Hernquist89}
{Hernquist}, L. \& {Katz}, N. 1989, \apjs, 70, 419

\bibitem[{{Hasegawa} \& {Semelin}(2013)}]{Hasegawa13}
{Hasegawa}, K., \& {Semelin}, B. 2013, \mnras, 428, 154

\bibitem[{{Hu} {et~al.}(2010){Hu}, {Cowie}, {Barger}, {Capak}, {Kakazu}, \&
  {Trouille}}]{Hu2010}
{Hu}, E.~M., {Cowie}, L.~L., {Barger}, A.~J., et al. 2010,
  \apj, 725, 394

\bibitem[{{Hui} \& {Gnedin}(1997)}]{Hui97}
{Hui}, L. \& {Gnedin}, N.~Y. 1997, \mnras, 292, 27

\bibitem[{{Iye} {et~al.}(2006){Iye}, {Ota}, {Kashikawa}, {Furusawa},
  {Hashimoto}, {Hattori}, {Matsuda}, {Morokuma}, {Ouchi}, \&
  {Shimasaku}}]{Iye06}
{Iye}, M., {Ota}, K., {Kashikawa}, N., et al. 2006,  \nat, 443, 186

\bibitem[{{Jeeson-Daniel} {et~al.}(2012){Jeeson-Daniel}, {Ciardi}, {Maio},
  {Pierleoni}, {Dijkstra}, \& {Maselli}}]{Jeeson-Daniel12}
{Jeeson-Daniel}, A., {Ciardi}, B., {Maio}, U., et al. 2012, \mnras, 424, 2193

\bibitem[{{Jeon} {et~al.}(2012){Jeon}, {Pawlik}, {Greif}, {Glover}, {Bromm},
  {Milosavljevic}, \& {Klessen}}]{Jeon2011}
{Jeon}, M., {Pawlik}, A.~H., {Greif}, T.~H., et al. 2012, \apj, 754, 34

\bibitem[{{Kashikawa} {et~al.}(2011){Kashikawa}, {Shimasaku}, {Matsuda},
  {Egami}, {Jiang}, {Nagao}, {Ouchi}, {Malkan}, {Hattori}, {Ota}, {Taniguchi},
  {Okamura}, {Ly}, {Iye}, {Furusawa}, {Shioya}, {Shibuya}, {Ishizaki}, \&
  {Toshikawa}}]{Kashikawa2011}
{Kashikawa}, N., {Shimasaku}, K., {Matsuda}, Y., et al. 2011,
  \apj, 734, 119

\bibitem[{{Katz} {et~al.}(2003){Katz}, {Keres}, {Dave}, \& {Weinberg}}]{Katz03}
{Katz}, N., {Keres}, D., {Dave}, R., \& {Weinberg}, D.~H. 2003, in Astrophysics
  and Space Science Library, Vol. 281, The IGM/Galaxy Connection. The
  Distribution of Baryons at z=0, ed. {J.~L.~Rosenberg \& M.~E.~Putman}, 185--+

\bibitem[{{Katz} {et~al.}(1996){Katz}, {Weinberg}, {Hernquist}, \&
  {Miralda-Escude}}]{Katz96}
{Katz}, N., {Weinberg}, D.~H., {Hernquist}, L., \& {Miralda-Escude}, J. 1996,
 \apjl, 457, L57+

\bibitem[{{Kennicutt}(1998)}]{Kennicutt98}
{Kennicutt}, Jr., R.~C. 1998, \araa, 36, 189

\bibitem[{{Kere{\v s}} {et~al.}(2009){Kere{\v s}}, {Katz}, {Fardal},
  {Dav{\'e}}, \& {Weinberg}}]{Keres09}
{Kere{\v s}}, D., {Katz}, N., {Fardal}, M., {Dav{\'e}}, R., \& {Weinberg},
  D.~H. 2009, \mnras, 395, 160

\bibitem[{{Kere{\v s}} {et~al.}(2005){Kere{\v s}}, {Katz}, {Weinberg}, \&
  {Dav{\'e}}}]{Keres05}
{Kere{\v s}}, D., {Katz}, N., {Weinberg}, D.~H., \& {Dav{\'e}}, R. 2005, \mnras, 363, 2

\bibitem[{{Komatsu} {et~al.}(2009){Komatsu}, {Dunkley}, {Nolta}, {Bennett},
  {Gold}, {Hinshaw}, {Jarosik}, {Larson}, {Limon}, {Page}, {Spergel},
  {Halpern}, {Hill}, {Kogut}, {Meyer}, {Tucker}, {Weiland}, {Wollack}, \&
  {Wright}}]{Komatsu09}
{Komatsu}, E., {Dunkley}, J., {Nolta}, M.~R., et al. 2009,
  \apjs, 180, 330

\bibitem[{{Latif} {et~al.}(2011){Latif}, {Zaroubi}, \& {Spaans}}]{Latif11}
{Latif}, M.~A., {Zaroubi}, S., \& {Spaans}, M. 2011, \mnras, 411,
  1659

\bibitem[{{Laursen} {et~al.}(2011){Laursen}, {Sommer-Larsen}, \&
  {Razoumov}}]{Laursen11}
{Laursen}, P., {Sommer-Larsen}, J., \& {Razoumov}, A.~O. 2011, \apj, 728, 52

\bibitem[{{Lehnert} {et~al.}(2010){Lehnert}, {Nesvadba}, {Cuby}, {Swinbank},
  {Morris}, {Cl{\'e}ment}, {Evans}, {Bremer}, \& {Basa}}]{Lehnert10}
{Lehnert}, M.~D., {Nesvadba}, N.~P.~H., {Cuby}, J., et al. 2010, \nat,
  467, 940

\bibitem[{{Li} {et~al.}(2007){Li}, {Hernquist}, {Robertson}, {Cox}, {Hopkins},
  {Springel}, {Gao}, {Di Matteo}, {Zentner}, {Jenkins}, \& {Yoshida}}]{Li07}
{Li}, Y., {Hernquist}, L., {Robertson}, B., et al. 2007, \apj, 665, 187

\bibitem[{{Li} {et~al.}(2008){Li}, {Hopkins}, {Hernquist}, {Finkbeiner}, {Cox},
  {Springel}, {Jiang}, {Fan}, \& {Yoshida}}]{Li08}
{Li}, Y., {Hopkins}, P.~F., {Hernquist}, L., et al. 2008, \apj, 678, 41

\bibitem[{{Malhotra} \& {Rhoads}(2004)}]{Malhotra04}
{Malhotra}, S. \& {Rhoads}, J.~E. 2004, \apjl, 617, L5

\bibitem[{{Matsuda} {et~al.}(2012)}]{Matsuda12}
{Matsuda}, Y., {Yamada}, T., {Hayashino}, T., et al. 2012, \mnras, 425, 878

\bibitem[{{Ocvirk} {et~al.}(2008){Ocvirk}, {Pichon}, \& {Teyssier}}]{Ocvirk08}
{Ocvirk}, P., {Pichon}, C., \& {Teyssier}, R. 2008, \mnras, 390, 1326

\bibitem[{{Ono} {et~al.}(2012){Ono}, {Ouchi}, {Mobasher}, {Dickinson},
  {Penner}, {Shimasaku}, {Weiner}, {Kartaltepe}, {Nakajima}, {Nayyeri},
  {Stern}, {Kashikawa}, \& {Spinrad}}]{Ono11}
{Ono}, Y., {Ouchi}, M., {Mobasher}, B., et al. 2012, \apj, 744, 83

\bibitem[{{Osterbrock} \& {Ferland}(2006)}]{Osterbrock06}
{Osterbrock}, D.~E. \& {Ferland}, G.~J. 2006, {Astrophysics of gaseous nebulae
  and active galactic nuclei}, ed. {Osterbrock, D.~E.~\& Ferland, G.~J.},
  {Astrophysics of gaseous nebulae and active galactic nuclei}

\bibitem[{{Ouchi} {et~al.}(2008){Ouchi}, {Shimasaku}, {Akiyama}, {Simpson},
  {Saito}, {Ueda}, {Furusawa}, {Sekiguchi}, {Yamada}, {Kodama}, {Kashikawa},
  {Okamura}, {Iye}, {Takata}, {Yoshida}, \& {Yoshida}}]{Ouchi08}
{Ouchi}, M., {Shimasaku}, K., {Akiyama}, M., et al. 2008, \apjs, 176, 301

\bibitem[{{Ouchi} {et~al.}(2010){Ouchi}, {Shimasaku}, {Furusawa}, {Saito},
  {Yoshida}, {Akiyama}, {Ono}, {Yamada}, {Ota}, {Kashikawa}, {Iye}, {Kodama},
  {Okamura}, {Simpson}, \& {Yoshida}}]{Ouchi10}
{Ouchi}, M., {Shimasaku}, K., {Furusawa}, H., et al. 2010,
 \apj, 723, 869

\bibitem[{{Rosdahl} \& {Blaizot}(2012)}]{Rosdahl12}
{Rosdahl}, J., \& {Blaizot}, J. 2012, \mnras, 423, 344

\bibitem[{Salpeter(1955)}]{Salpeter55}
Salpeter, E.~E. 1955, ApJ, 121,
  161

\bibitem[{{Santos}(2004)}]{Santos04}
{Santos}, M.~R. 2004, \mnras, 349, 1137

\bibitem[{{Scannapieco} {et~al.}(2012){Scannapieco}, {Wadepuhl}, {Parry},
  {Navarro}, {Jenkins}, {Springel}, {Teyssier}, {Carlson}, {Couchman}, {Crain},
  {Dalla Vecchia}, {Frenk}, {Kobayashi}, {Monaco}, {Murante}, {Okamoto},
  {Quinn}, {Schaye}, {Stinson}, {Theuns}, {Wadsley}, {White}, \&
  {Woods}}]{Scannapieco12}
{Scannapieco}, C., {Wadepuhl}, M., {Parry}, O.~H., et al. 2012, \mnras, 423, 1726

\bibitem[{{Schmidt}(1959)}]{Schmidt59}
{Schmidt}, M. 1959, \apj, 129, 243

\bibitem[{{Shibuya} {et~al.}(2012){Shibuya}, {Kashikawa}, {Ota}, {Iye},
  {Ouchi}, {Furusawa}, {Shimasaku}, \& {Hattori}}]{Shibuya11}
{Shibuya}, T., {Kashikawa}, N., {Ota}, K., et al. 2012, \apj, 752, 114

\bibitem[{{Springel}(2005)}]{Springel05e}
{Springel}, V. 2005, MNRAS, 364,
  1105

\bibitem[{{Springel} {et~al.}(2005){Springel}, {Di Matteo}, \&
  {Hernquist}}]{Springel05d}
{Springel}, V., {Di Matteo}, T., \& {Hernquist}, L. 2005, MNRAS, 361, 776

\bibitem[{{Springel} \& {Hernquist}(2002)}]{Springel02}
{Springel}, V. \& {Hernquist}, L. 2002, MNRAS, 333, 649

\bibitem[{{Springel} \& {Hernquist}(2003)}]{Springel03a}
---. 2003, MNRAS, 339, 289

\bibitem[{{Springel} {et~al.}(2008){Springel}, {Wang}, {Vogelsberger},
  {Ludlow}, {Jenkins}, {Helmi}, {Navarro}, {Frenk}, \& {White}}]{Springel08a}
{Springel}, V., {Wang}, J., {Vogelsberger}, M., et al. 2008,
   MNRAS, 391, 1685

\bibitem[{{Springel} {et~al.}(2001){Springel}, {Yoshida}, \&
  {White}}]{Springel01}
{Springel}, V., {Yoshida}, N., \& {White}, S.~D.~M. 2001, New Astronomy, 6,
  79

\bibitem[{{Stark} {et~al.}(2011){Stark}, {Ellis}, \& {Ouchi}}]{Stark2011}
{Stark}, D.~P., {Ellis}, R.~S., \& {Ouchi}, M. 2011, \apjl, 728, L2

\bibitem[{{Stark} {et~al.}(2007){Stark}, {Ellis}, {Richard}, {Kneib}, {Smith},
  \& {Santos}}]{Stark07}
{Stark}, D.~P., {Ellis}, R.~S., {Richard}, J., et al. 2007, \apj, 663, 10

\bibitem[{{Thoul} \& {Weinberg}(1996)}]{Thoul96}
{Thoul}, A.~A., \& {Weinberg}, D.~H. 1996, \apj, 465, 608

\bibitem[{{Vanzella} {et~al.}(2011){Vanzella}, {Pentericci}, {Fontana},
  {Grazian}, {Castellano}, {Boutsia}, {Cristiani}, {Dickinson}, {Gallozzi},
  {Giallongo}, {Giavalisco}, {Maiolino}, {Moorwood}, {Paris}, \&
  {Santini}}]{Vanzella11}
{Vanzella}, E., {Pentericci}, L., {Fontana}, A., et al. 2011, \apjl, 730, L35

\bibitem[{{Wadepuhl} \& {Springel}(2011)}]{Wadepuhl11}
{Wadepuhl}, M. \& {Springel}, V. 2011, \mnras, 410, 1975

\bibitem[{{Wise} \& {Abel}(2007)}]{Wise2007}
{Wise}, J.~H. \& {Abel}, T. 2007, \apj, 665, 899

\bibitem[{{Wise} \& {Abel}(2008)}]{Wise2008B}
---. 2008, \apj, 685, 40

\bibitem[{{Wise} {et~al.}(2008){Wise}, {Turk}, \& {Abel}}]{Wise2008A}
{Wise}, J.~H., {Turk}, M.~J., \& {Abel}, T. 2008, \apj, 682, 745

\bibitem[{{Wise} {et~al.}(2012){Wise}, {Turk}, {Norman}, \& {Abel}}]{Wise2010}
{Wise}, J.~H., {Turk}, M.~J., {Norman}, M.~L., \& {Abel}, T. 2012,  \apj, 745, 50

\bibitem[{{Yajima} {et~al.}(2012{\natexlab{a}}){Yajima}, {Li}, {Zhu}, \&
  {Abel}}]{Yajima12b}
{Yajima}, H., {Li}, Y., {Zhu}, Q., \& {Abel}, T. 2012{\natexlab{a}},
 \mnras, 424, 884

\bibitem[{{Yajima} {et~al.}(2012{\natexlab{b}}){Yajima}, {Li}, {Zhu}, {Abel},
  {Gronwall}, \& {Ciardullo}}]{Yajima12c}
{Yajima}, H., {Li}, Y., {Zhu}, Q., et al. 2012{\natexlab{b}}, \apj, 754, 118

\bibitem[{{Yajima} {et~al.}(2012{\natexlab{c}}){Yajima}, {Li}, {Zhu}, {Abel},
  {Gronwall}, \& {Ciardullo}}]{Yajima12d}
{Yajima}, H., {Li}, Y., {Zhu}, Q., et al. 2012{\natexlab{c}}, submitted to \mnras, arXiv: 1209.5842

\bibitem[{{Yoshida} {et~al.}(2007){Yoshida}, {Oh}, {Kitayama}, \&
  {Hernquist}}]{Yoshida07}
{Yoshida}, N., {Oh}, S.~P., {Kitayama}, T., \& {Hernquist}, L. 2007, \apj, 663, 687

\bibitem[{{Yoshida} {et~al.}(2008){Yoshida}, {Omukai}, \&
  {Hernquist}}]{Yoshida2008}
{Yoshida}, N., {Omukai}, K., \& {Hernquist}, L. 2008, Science, 321, 669

\bibitem[{{Zheng} {et~al.}(2010){Zheng}, {Cen}, {Trac}, \&
  {Miralda-Escud{\'e}}}]{Zheng10}
{Zheng}, Z., {Cen}, R., {Trac}, H., \& {Miralda-Escud{\'e}}, J. 2010,
  \apj, 716, 574

\bibitem[{{Zheng} \& {Miralda-Escud{\'e}}(2002)}]{Zheng02}
{Zheng}, Z. \& {Miralda-Escud{\'e}}, J. 2002, \apj, 578, 33

\bibitem[{{Zhu} {et~al.}(2010)}]{Zhu12}
{Zhu}, Q. and {Li}, Y. and {Sherman}, S. 2012, submitted to \apj, arXiv: 1211.0013

\end{thebibliography}

\end{document}